\documentclass[12pt]{aastex}
\def \lp{\>\> .}
\def \lc{\>\> ,}

\def \c2{cm$^{-2}$}
\def \cc{cm$^{-3}$}

\def \kms{km\ s$^{-1}$}

\def \nh2{n_{H_2}}
\def \nh1{n_{HI}}

\def \nh3{NH$_3$}
\def \n2h{N$_2$H$^+$}
\def \tw{$^{12}$CO}
\def \th{$^{13}$CO}

\def\ho{H$^0$}
\def \hh{H$_2$}

\def \ox18{$^{16}$O$^{18}$O}

\def \C2{C{\sc II}}
\def\C1{C{\sc I}}
\def\O1{O{\sc I}}
\def \H1{H{\scI}}
\def\h0{H$^0$}

\def\n3{$n_u$}
\def\n1{$n_l$}

\def \mic{$\mu$m}

\def \be{\begin{equation}}
\def \ee{   \end{equation}}
\def \bf {\begin{figure}}
\def \ef {   \end{figure}}
\def \bc{\begin{center}}
\def \ec{\begin{center}}

\def \lc{\>\> ,}
\def \lp{\>\> .}

\onecolumn
\begin{document}
\title{Diffuse Molecular Cloud Densities from UV Measurements of CO Absorption}                        

\setcounter{footnote}{0}

\author{Paul F. Goldsmith}
\affil{Jet Propulsion Laboratory, California Institute of Technology}

\section{Abstract}

We use UV measurements of interstellar CO towards nearby stars to calculate the density in the diffuse molecular clouds containing the molecules responsible for the observed absorption.  
Chemical models and recent calculations of the excitation rate coefficients indicate that the regions in which CO is found have hydrogen predominantly in molecular form.   
We carry out statistical equilibrium calculations using CO--\hh\ collision rates to solve for the \hh\ density in the observed sources without including effects of radiative trapping.
We have  assumed kinetic temperatures of 50 K and 100 K, finding this choice to make relatively little difference to the lowest transition.  
For the sources having $T^{ex}_{10}$ only, for which we could determine upper and lower density limits, we find $<n(H_2)>$ = 49 \cc.  
While we can find a consistent density range for a good fraction of the sources having either two or three values of the excitation temperature, there is a suggestion that the higher--$J$ transitions are sampling clouds or regions within diffuse molecular cloud material that have higher densities than the material sampled by the $J$ = 1--0 transition.
The assumed kinetic temperature and derived \hh\ density are anticorrelated when the $J$ = 2--1 transition data, the J = 3--2 transition data, or both are included.
For sources with either two or three values of the excitation temperature, we find average values of the midpoint of the density range that is consistent with all of the observations equal to 68 \cc\ for $T^k$ = 100 K and 92 \cc\ for $T^k$ = 50 K. 
The data for this set of  sources imply that diffuse molecular clouds are characterized by an average thermal pressure between 4600 and 6800 K\cc.

Keywords:  ISM:  molecules - radio lines:  ISM

\section{Introduction}

Diffuse clouds have been studied over a broad range of wavelengths encompassing radio observations of the 21 cm H\,{\sc i} line to UV observations of \hh\ and other molecules.  
They have been been found to encompass a wide range of densities, temperatures, and column densities.
For low column densities, the gas is essentially atomic (H$^0$) and ionic (C$^+$), but as the column density and extinction increase, molecules (starting with \hh) gradually become dominant, and the term ``diffuse molecular cloud'' \citep{snow2006} is appropriate.  
Ground--based observations of strong millimeter continuum sources \citep{liszt1998} and UV observations of early--type stars \citep[e.g.][]{sheffer2008} have both allowed observations of rotational transitions of the CO molecule.
The UV observations are particularly powerful in that they allow simultaneous measurements of multiple transitions, which are sensitive to the column density of the different CO rotational levels in the cloud in the foreground of the star.  
The column density of \hh\ can also be determined, which allowing determination of the abundances of a number of different species and also isotope ratios as a function of column density \citep{sheffer2007}.

The density of diffuse molecular clouds is an important parameter for analyzing emission in various tracers as well as determining a number of critical cloud properties such as their thermal pressure.
One important tracer of several phases of the interstellar medium including diffuse clouds is the fine structure transition of ionized carbon      ([C\,{\sc ii}]).  Several groups using the  Herschel Satellite \citep{pilbratt2010} have carried out extensive observations of this submillimeter ($\lambda$ = 158 \mic) transition.  
Among these, a large-scale survey of the Milky Way has been attempting to apportion the observed [CII] emission among the different phases of the interstellar medium \citep{langer2010, pineda2013}.  
The [C\,{\sc ii}] emission from the diffuse cloud component of the interstellar medium will almost certainly be subthermal given that the critical density for the  [C\,{\sc ii}] fine structure line  is $\simeq$ 2000 -- 6000 \cc\ \citep{goldsmith2012}.  
Since the [C\,{\sc ii}] transition is optically thin or in the effectively optically thin limit \citep{goldsmith2012}, the inferred column density of ionized carbon in the diffuse interstellar medium will vary inversely as the density in the clouds responsible for the [CII] emission.
The cooling and thermal balance are also sensitively dependent on the density, so that understanding the structure of diffuse clouds and their role in the formation of denser clouds and star formation requires knowledge of the density in the diffuse interstellar medium.

In this paper we use the relative populations of the lower CO rotational levels to determine the density in the diffuse clouds  along the line of sight to early--type stars observed in the UV.  
Data on sixty four sources were taken from \citet{sheffer2008}.  
We have supplemented these data with observational results on eight distinct sources observed by \citet{sonnentrucker2007}, who also present data obtained by \citet{lambert1994} and \citet{federman2003} for three sources.  
Two additional, distinct sources were observed by \citet{burgh2007}.  

In Section \ref{excitation_temperatures} we discuss the transformation of the \citet{sheffer2008}  data to standard excitation temperatures that characterize successive rotational transitions, and in Section \ref{uncertainties} derive the uncertainties in the excitation temperatures resulting from their column density measurements.
In section \ref{co_excitation} we discuss the possibility of radiative excitation, and conclude that it is unlikely to play a significant role.  
We focus on collisional excitation of CO, concluding that collisions with \hh\ molecules are dominant in the clouds of interest.
Section \ref{densities} gives the results for different categories of diffuse clouds defined by which CO transitions have been observed.  
In Section \ref{discussion} we discuss and summarize our results.
The Appendix gives an explanation of long--standing apparently anomalous results for the excitation temperature in the low--density limit from multilevel statistical equilibrium calculations that can, in fact, be understood in terms of the allowed collisions and the spontaneous decay rates.

\section{Excitation Temperatures}
\label{excitation_temperatures}

The excitation temperature, $T^{ex}$,  is defined by the relative local densities in two different energy levels, or (having a clear physical meaning if conditions are uniform along the line of sight) by the relative column densities, $N$, of the two levels of a given species.  
Denoting the upper and lower levels by $u$ and $l$, respectively, and their statistical weights by $g_u$ and $g_l$, the relationship is
\be
\label{nunl}
\frac{N_u}{N_l} = \frac{g_u}{g_l}\exp[-\Delta E_{ul}/kT^{ex}_{ul}] \lc
\ee
where $\Delta E_{ul}$ is the energy difference between the upper and the lower level.
The excitation temperature can be defined between any pair of levels, but it is of greatest utility for two levels connected by a radiative transition that can be observed.

\citet{sheffer2008} use UV absorption measurements to determine the column densities in a number of the lowest transitions of the carbon monoxide (CO) molecule, and define the excitation temperatures of the excited rotational levels ($J$ = 1, 2, 3) relative to the ground state, $J$ = 0.  
The lowest excitation temperature thus defined corresponds to the CO $J$ = 1 -- 0 transition at 115.3 GHz.  
The excitation temperatures related to column densities of the $J$ = 2 and $J$ = 3 levels relative to $J$ = 0 do not correspond to observable transitions.
It is convenient for density determinations to deal with pairs of levels connected by a radiative transition, so that the collision rate directly competes with an allowed radiative processes.
The results tabulated by \citet{sheffer2008} can easily be transformed into the desired excitation temperatures through the following relationships, in which $T_{01}$, $T_{02}$, and $T_{03}$ are the excitation temperatures of the indicated pairs of levels determined by \citet{sheffer2008}, and $T^{ex}_{01}$, $T^{ex}_{21}$, and $T^{ex}_{32}$ are the excitation temperatures for radiatively--connected pairs of levels.  
We define for each transition the equivalent temperature, $T^*_{ul}$ = $\Delta E_{ul}/k$,

\begin{eqnarray}
T^{ex}_{1 0} &=& T_{01} \\
T^{ex}_{2 1} &=& T^*_{2 1}\frac{T_{0 2}T_{01}}{T^*_{2 0}T_{01} - T^*_{1 0}T_{0 2}} \\
T^{ex}_{3 2} &=& T^*_{3 2}\frac{T_{0 3}T_{02}}{T^*_{3 0}T_{02} - T^*_{2 0}T_{0 3}} \lp
\end{eqnarray}

The transformed results for the stars observed by \citet{sheffer2008} are given in Table \ref{texdata}, along with the molecular hydrogen column density determined for each line of sight.
In two cases, the \hh\ column density was estimated by \citet{sheffer2008} from the column densities of CO and CH, and these values are singled out by a note in the Table.
For four lines of sight, \citet{sheffer2008} did not include N(\hh), but values for these were found in the literature and values with associated references are given in column 5 of Table \ref{texdata}.
The data in Table 6 of \cite{lambert1994}, Table 1 of \citet{burgh2007}, and Table 12 of \citet{sonnentrucker2007} are presented in the form of excitation temperatures between adjacent rotational levels and so can be used directly.
These data are presented in Table \ref{others}, along with references to the original observational papers.

\section{Uncertainties}
\label{uncertainties}

In deriving excitation temperatures, $T_{ex}$ from column densities $N$, we use the usual relationship
given in equation \ref{nunl}, which leads to the expression for the excitation temperature
\be
\label{texdef}
T^{ex} = \frac{T^*}{ln[\frac{N_l}{N_u} \frac{g_u}{g_l}]} \lp
\ee

Taking the partial derivatives with respect to upper and lower level column densities, we find
\be
\frac{dT^{ex}}{T^{ex}} = \frac{T^{ex}}{T^*}[\frac{dN_l}{N_l} - \frac{dN_u}{N_u}] \lp
\ee
Defining the rms uncertainties as $\sigma_{T^{ex}}$, $\sigma_{N_l}$, and $\sigma_{N_u}$, respectively, and combining the fractional uncertainties  as the sum of the squared uncertainties in the lower and upper level column densities gives us
\be
\frac{\sigma_{T^{ex}}}{T^{ex}} = \frac{T^{ex}}{T^*}[(\frac{\sigma_{N_l}}{N_l})^2 + (\frac{\sigma_{N_u}}{N_u})^2]^{0.5} \lp
\ee

For the UV absorption data of interest, the excitation temperatures are on the order of 0.6 to 0.8 times $T^*$ (e.g. 4 K for the $J$ = 1-0 transition having $T^*$ = 5.5 K). 
It is thus reasonable to take $T^{ex}/T^*$ $\simeq$ 0.7, which gives us
\be
\frac{\sigma_{T^{ex}}}{T^{ex}} = 0.7[(\frac{\sigma_{N_l}}{N_l})^2 + (\frac{\sigma_{N_u}}{N_u})^2]^{0.5} \lp
\ee
 
 Sheffer et al. (2008) give only the uncertainty in the total column density of CO.  
 While it is not clear exactly how the uncertainty in the total column density is related to the fractional uncertainty in the column density of a single level, we simply assume that the fractional uncertainty in an individual column density is equal to the total CO column density uncertainty, given as 20\% by \cite{sheffer2008}.
 Then the fractional uncertainty in the excitation temperature is $\simeq$ 0.7$\sqrt2$ times 20\%.  
 It thus seems that a reasonably generous  1$\sigma$ value is $\sigma_{T^{ex}}/T^{ex}$ = 0.2.  
 
 The observations taken from other papers (Table \ref{others}) explicitly include uncertainties in individual excitation temperatures.  
 As seen in that Table, these vary considerably from source to source, but are of the same order as given by the above analysis.
 
 \section{CO Excitation}
 \label{co_excitation}
 
 \subsection{Non--Collisional Excitation}

The excitation of CO can, in principle, be affected by radiative processes following its formation.
The unshielded photodissociation rate of $^{12}$CO in a radiation field having the standard Draine value \citep{draine1978} is $k_{i0} = 2\times10^{-10}$ s$^{-1}$ \citep{umist2012}. 
For the \hh\ and CO column densities of the clouds in this sample, the shielding factor is $\simeq$ 0.5  \cite[see][Table 5]{vandishoeck1988}, and thus the CO photodissociation rate within the cloud, which we take equal to the formation rate, is on the order of 10$^{-10}$ s$^{-1}$.  
The characteristic time scale is thus $\simeq$ 300 yr.  
The vibrational decay rate is enormously faster, with $A(v=1 - v=0)$ = 30.6 s$^{-1}$ \citep{chandra1996}.
Thus, any CO molecule formed will very rapidly decay to the ground vibrational state.
The spontaneous decay rates for the rotational transitions are many orders of magnitude slower, ranging from $A_{10}$ = 7.2$\times$10$^{-8}$ s$^{-1}$ to $A_{32}$ = 2.5$\times$10$^{-6}$ s$^{-1}$ for the transitions considered here \citep{cdms2013}.

The collision rates necessary to achieve the observed subthermal excitation of CO (see Section \ref{densities}) are $\simeq$ 10$^{-8}$ s$^{-1}$ or 100 times the formation timescale.
Thus with all CO molecules being in the ground vibrational state, the collision rate that determines the rotational level populations will be much more rapid than formation/destruction rate, and it is reasonable that the effect of a post--formation cascade (as can affect the population of the levels of \hh) will be unimportant.

\citet{wannier1997} suggested that the radiation from a nearby, dense molecular cloud could be sufficient to provide the observed excitation of CO in a diffuse molecular cloud.  
This requires that the two clouds have the same velocity and that the solid angle of the cloud providing the radiative pumping be large enough to make the radiative excitation rate comparable to the spontaneous decay rate of the transition observed.  
\citet {sonnentrucker2007} pointed out that a critical test of this model follows from the fact that the pumping cloud, while optically thick in \tw\  would almost certainly be optically thin in \th.  
The result would be a much lower radiative pumping rate for \th\ than for \tw, and the excitation temperatures of the rare isotopologue would thus be significantly smaller.  
\citet{sonnentrucker2007} conclude that for 7 sight lines (including some observed by others) $T^{ex}_{10}$(\tw) is, within the uncertainties, equal to that of \th.  

An additional consideration is that the excitation temperature $T^{ex}_{J, J-1}$(\tw) increases with increasing $J$; this increase is predicted by the collisional excitation model discussed in the Appendix.
For \th, there are only two sources with excitation temperatures determined for more than one transition.
Both of these, HD147933 \citep{lambert1994} and HD24534 \citep{sonnentrucker2007}, show this behavior.
Given the constraints imposed by the limited signal to noise ratio, it is difficult to be definitive, but we agree with \cite{sonnentrucker2007} that radiative excitation by nearby clouds does not play a major role in determining the excitation temperature of the lower rotational transitions of CO and that excitation is primarily by collisions.
\citet{zsargo2003} similarly concluded that optical pumping is generally unimportant for excitation of CI in diffuse clouds.

\subsection{Collisional Excitation of CO in Diffuse Molecular Clouds}
\label{partners}

Analyzing the excitation of CO and determining the density of diffuse clouds is linked to their structure.
Possibly important collision partners in diffuse clouds are electrons, atomic hydrogen (H$^0$) and molecular hydrogen (H$_2$).  In these clouds, carbon is largely in ionized form (see, for example Figure \ref{av1n100model}) so the fractional abundance of electrons $\simeq$ 10$^{-4}$ throughout diffuse molecular clouds.  
\citet{crawford1971} calculated the cross sections for excitation of low--$J$ transitions of CO due to collisions with electrons. 
Their results are reproduced by the general, but more approximate treatment of \citet{dickinson1975}, who calculate excitation rate coefficients and fit a quite convenient general formula.
Application to the low--dipole moment CO molecule, yields characteristic deexcitation rate coefficients between 0.4 and 0.5 $\times$10$^{-8}$ cm$^{3}$s$^{-1}$, for kinetic temperatures between 50 K and 150 K, and for $J_{upper}$ between 1 and 6.
These are approximately a factor of 100 larger than those for collisions with atomic or molecular hydrogen (see discussion in Section \ref{H_H2_rates}).  

However, this is not a sufficient factor to compensate for the much lower fractional abundance of electrons, and in consequence electrons should not be a significant source of collisional excitation for CO in diffuse clouds.
The more detailed discussion in Section \ref{H_H2_rates} thus considers only excitation by collisions with H$^0$ and H$_2$.
Note that this situation is quite different than that for high-dipole moment molecules such as HCN \citep{dickinson1977}, since the  coefficients for electron excitation scale approximately as $\mu^2$.
Thus for HCN or CN, electron excitation rate coefficients will be 10$^4$ to 10$^5$ times larger than those for collisions with atoms or molecules.

\subsection{Cloud Structure, the H$^0$ to H$_2$ Transition, and Excitation Analysis}

We are left with atomic and molecular hydrogen as being significant for collisional excitation of CO in diffuse molecular clouds.  
The distribution and abundance of each varies through a cloud due to the competition between formation and photodissociation; the latter is mediated by self-shielding.
The processes determining the transformation between \h0\ and \hh\ are well-treated by the Meudon PDR code \citep{lepetit2006}.
We have carried out a number of runs modeling slabs exposed to the interstellar radiation field on both sides, with a uniform density defined by $n(\rm H)$ = $n$(\h0) + 2$n$(\hh).  
The critical results are summarized in Table \ref{pdrmodels}.
We include the molecular fraction defined by $f$(\hh) = 2$n$(\hh)/(2$n$(\hh) + $n$(\h0)), defined in the central portion of the cloud, and also integrated through the entire cloud, which we denote $F$(\hh) following \citet{snow2006}.  
For clouds having extinction exceeding a few tenths of a magnitude, a large fraction of hydrogen is in molecular form.  
What is particularly important to note is that the \hh\ fraction in the centers of the slab is high, generally $\geq$ 0.75, and in some  relevant cases, $>$ 0.9.
 
From the data on color excess presented by \cite{rachford2002} and \cite{sheffer2008} we can determine the total hydrogen column density $N(\rm H)$ and the integrated hydrogen fraction for some of the sources observed here.
The values for most sources are $\simeq$ 0.5, confirming that, for the sources here, a large fraction of the hydrogen will be in molecular form.
This is consistent with the information presented in Table 2 of \citet{burgh2007} showing that $F$(\hh) $\geq$ 0.24 for 8 of the 9 sources with $N$(\hh) $>$ 10$^{20}$ cm$^{-2}$.  
The one exception, HD102065, has a reasonable density range determined with a single CO transition (Table \ref{dens_1-0_only}), but an enhanced UV field could result in the low integrated molecular fraction, $F$(\hh) = 0.1 \citep{burgh2007}.
$F$(\hh) is not obviously correlated with $N(\rm H)$, suggesting that other characteristics such as cloud density and environment play an important role in determining the balance between atomic and molecular hydrogen.   

This situation is illustrated by the cloud model results shown in Figure \ref{av1n100model}.
The \hh\ density exceeds that of \ho\ for visual extinctions $\geq$ 0.03 mag, and in the central region of the cloud, $n$(\hh) $\simeq$ 50 \cc, which is quite similar to the average value determined below in Section \ref{densities}.
The kinetic temperature varies between $\simeq$ 50 K and $\simeq$ 100 K throughout the cloud, also in good agreement with observations (e.g. Table 6 of Sheffer et al. 2008).
While this treatment does not consider all combinations of extreme conditions, it is reasonable to conclude that diffuse molecular clouds have a largely molecular hydrogen core, surrounded by a region in which the hydrogen is primarily atomic.
The size of the molecular core, the peak \hh\ fraction, and the integrated \hh\ fraction all increase with increasing cloud density, and decrease as the strength of the interstellar radiation field increases.  
It is thus plausible that in diffuse clouds with visual extinction of a few tenths to $\simeq$ 1 mag, such as most of those observed in the above--cited papers, the density of \hh\ is a factor 2 to 10 times larger than that of \h0.  
This is consistent with the properties suggested for ``transitional clouds'' studied in HI self--absorption by \citet{kavars2005}.

A consideration when comparing these models with observations is the question of multiple clouds along the line of sight.
\citet{welty2001} find that their extremely high spectral resolution (ground--based) observations require multiple, relatively narrow velocity components in order to obtain good fits to their observed K{\sc I} line profiles.  
Such resolution is not available for UV observations of CO, but the observed cloud parameters may, in fact, refer to the sum of a number of individual components.  

The major effect of multiple clouds is that the extinction in each individual component cloud is smaller than the total line of sight extinction. 
The clouds being considered here have (measured) \hh\ column densities between 1 and 6 $\times$ 10$^{20}$ \c2.
This alone corresponds to extinctions between 0.1 and 0.6 mag.
If we assume a nominal integrated \hh\ fraction $F$(\hh) = 0.6, $N$(\ho) = 1.33$N$(\hh), and the atomic hydrogen column density envelope raises the total extinction through the cloud to $\simeq$ 0.2 -- 1.0 mag.  
If we have, for example, three equal component clouds along the line of sight, each has extinction between 0.07 and 0.33 mag.  
In conditions of standard radiation field intensity and $n$(H) $\geq$ 100 \cc, the peak \hh\ fraction will reach 0.5 for the lowest column density clouds, and will be close to unity for those having the highest column densities.
Thus, even the presence of a modest number of components along the line of sight will not change the basic picture of a constituent cloud having a primarily molecular \hh\ core surrounded by a \h0 envelope.

%%%%%%%FIGURE 1
\bf
\begin{center}
\includegraphics[width = 0.8\textwidth]{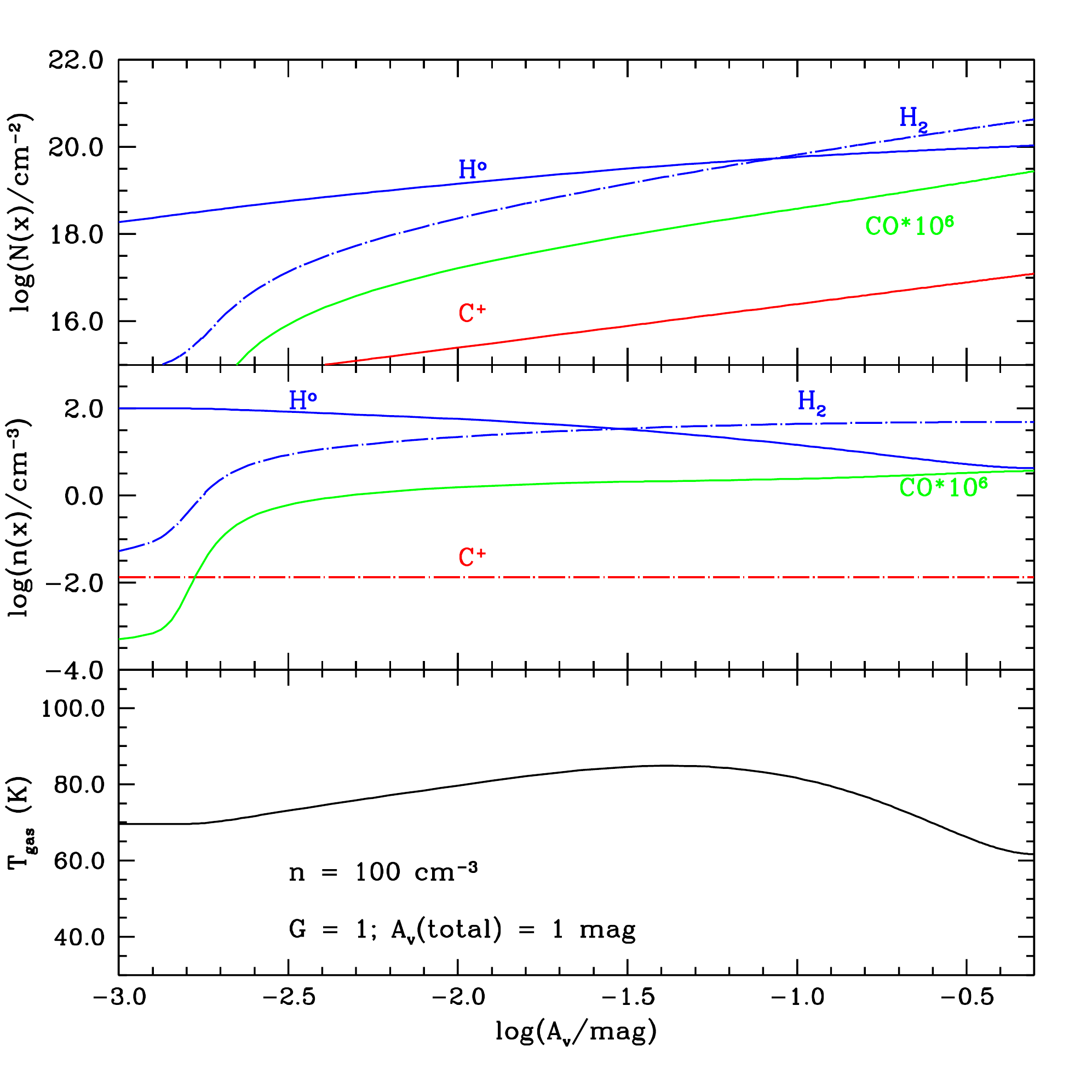}
\caption{\label{av1n100model}  Results of model calculation using the Meudon PDR code for a two--sided slab exposed on both sides to an interstellar radiation field of standard intensity.  The proton density $n(\rm H)$ is equal to 100 \cc.  The bottom panel shows the variation of the kinetic temperature through the cloud.   The middle panel shows the volume densities of ionized carbon, CO, \h0, and \hh. 
Note that the \hh\ density in the central portion of the cloud is close to 50 \cc. The upper panel shows the column densities of these four species integrated from the edge of the cloud. 
}
\end{center}
\ef
%%%%%%%%%%%%

\subsection{CO Chemistry}

Another factor is the chemistry of CO. 
While it is not appropriate to go into this in much detail here, a short review is important to appreciate where in the observed clouds the CO being observed is actually located.
In clouds where hydrogen is atomic, the only route to form CO starts with the radiative association reaction of C$^+$ and \h0.  
This reaction is extremely slow, but the CH$^+$ that forms yields (through reaction with O) a slow rate of CO production, and combined with relatively unattenuated photodissociation, results in a low fractional abundance of CO.
In clouds with the hydrogen in the form of \hh, a similar radiative association reaction between C$^+$ (which will still be the dominant form of carbon due to its lower ionization potential) and \hh\ can take place, but it is $\simeq$ 40 times faster than that with \h0, leading to somewhat higher CO fractional abundances.
A second path is the chemical reaction C$^+$ + \hh\ $\rightarrow$ CH$^+$ + H.
However, this reaction is endothemic by 4640 K, and thus is extremely slow at normal cloud temperatures.  
The above pathways lead to a fractional abundance of CO in diffuse molecular regions $\simeq$ 3$\times$10$^{-8}$, similar to that seen in Figure \ref{av1n100model}, but significantly below that observed for diffuse molecular clouds being considered here ($\langle${N({\rm CO})/N({\rm H$_2$}$\rangle$ = 3$\times$10$^{-7}$; \citet{federman1980}, \citet{sheffer2008}).

\citet{elitzur1978} suggested that presence of shock heating would significantly raise the temperature of the molecular gas and enhance the abundance of CH$^+$.  
This would also have the effect of increasing the abundance of CO. 
This could resolve the discrepancy between model and observations, but has the undesirable consequence of copiously producing OH via the reaction O + \hh\ $\rightarrow$ OH + H, which is endothermic by 3260 K.  
The predicted OH fractional abundance exceeds that observed by a large factor.

In order to exploit the rapid reaction between C$^+$ and \hh\ at high temperatures without overproducing OH, \citet{federman1996} suggested that Alfv\'{e}n waves could heat diffuse clouds and the outer portions of larger molecular clouds with the special effect of raising the temperature of the ions and not that of the neutrals.  
Thus, CH$^+$, and CO abundances could be enhanced without overproducing OH.  
This ''superthermal" chemistry was supported by observations of various species by \citet{zsargo2003} and has subsequently been incorporated into different models, notably that of \citet{visser2009}, that successfully reproduce the run of CO {\it vs} \hh\ column densities.

An alternative explanation that explains the abundances of a number of species in diffuse molecular clouds is heating in regions of turbulent dissipation, discussed by \citet{godard2009}.
What is essential for the present discussion is that these models are entirely dependent on having molecular hydrogen as the starting point. 
In contrast, no models starting with atomic hydrogen can achieve fractional abundances of CO close to those observed.  
Thus, the chemistry of CO strongly suggests that we are tracing a species confined to the portion of the cloud in which hydrogen is largely in the form of \hh.
Note that the Meudon PDR code does not include superthermal chemistry so that the CO fractional abundances predicted (e.g. Figure \ref{av1n100model}) are significantly below those derived from observations.

\subsection{Collision Rate Coefficients}
\label{H_H2_rates}

Due to the importance of CO in the dense interstellar medium in which hydrogen is almost entirely molecular, cross sections and excitation rates for collisions between CO and \hh\ have been calculated by several different groups starting more than three decades ago \citep{green1976,flower2001,wernli2006,yang2010}.  
The earliest calculations treated \hh\ molecules as He atoms (with a simple scaling for their different mass), which is correct only for the lowest, spherically symmetric level (J = 0) of para--\hh.  
This calculation is thus strictly speaking not applicable to ortho--\hh, which is expected to have comparable or even greater abundance than the para--\hh\ spin modification.
While the details of the potential surfaces and the quantum calculations have evolved, the results are all quite similar.
Figure \ref{xsec_comp}  includes the more detailed calculations that treat ortho--\hh\ and para--\hh\ as separate species.  
As seen in Figure \ref{xsec_comp}, there is very little difference between the deexcitation rate coefficients for ortho-- and for para-- \hh.  

%%%%%%%FIGURE 2  %%%%%%%%%%%%%%%%%%%%%%%%%%%%%%
\bf
\begin{center}
\includegraphics[width = 0.8\textwidth]{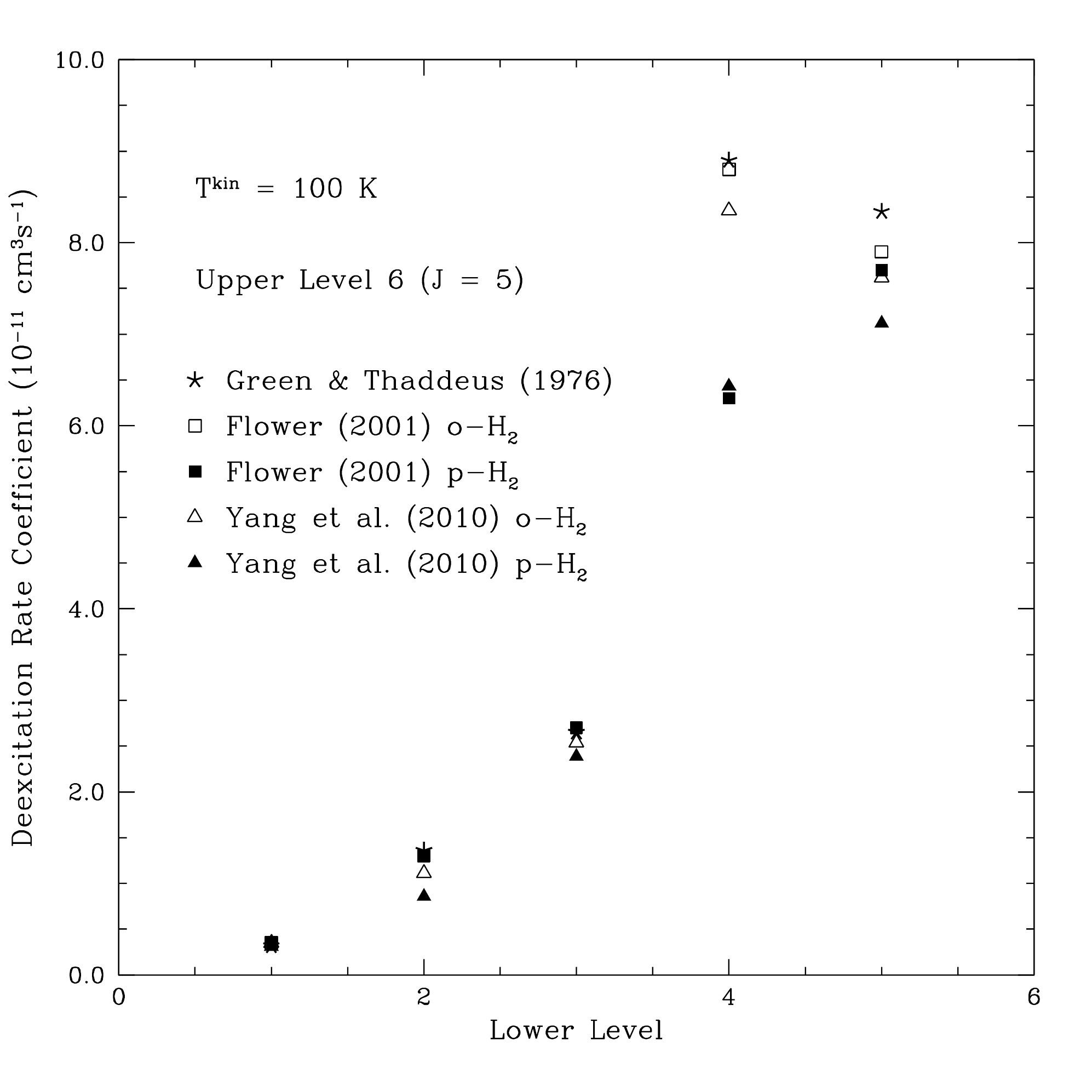}
\caption{\label{xsec_comp}  Comparison of different calculations of the CO deexcitation rate coefficients for collisions with \hh.  This limited sample, which includes only initial level 6 (J = 5) transitions to lower levels, indicates agreement to within $\pm$ 20\%.  
}
\end{center}
\ef
%%%%%%%%%%%%%%%%%%%%%%%%%%%%%%%%%%%%%%%%%%

To model the excitation of CO by collisions with \hh\ in diffuse clouds, we use the most recent results of \citet{yang2010} as given in the LAMDA database.
We have adopted an ortho--to--para \hh\ ratio (OPR) equal to 3, but varying the OPR from 1 to 3 does not make a significant difference in the excitation temperature for a given total \hh\ density.
The results for the three lowest CO transitions for three kinetic temperatures representative of the temperatures measured for this sample of diffuse molecular cloud sight lines are shown in Figure \ref{tex_all_tk}.
We see that the excitation temperatures of all three transitions increase monotonically with the \hh\ density throughout this range.
For densities below $\simeq$ 50 \cc, the excitation temperatures $T^{ex}_{JJ-1}$ increase as $J$ increases.  
This apparently surprising behavior is discussed in detail in the Appendix.
All transitions are assumed to be optically thin for these calculations.

%%%%%%%%%FIGURE 3  %%%%%%%%%%%%%%%%%%%%%%%%%%%%%
\bf
\begin{center}
\includegraphics[width = 0.8\textwidth]{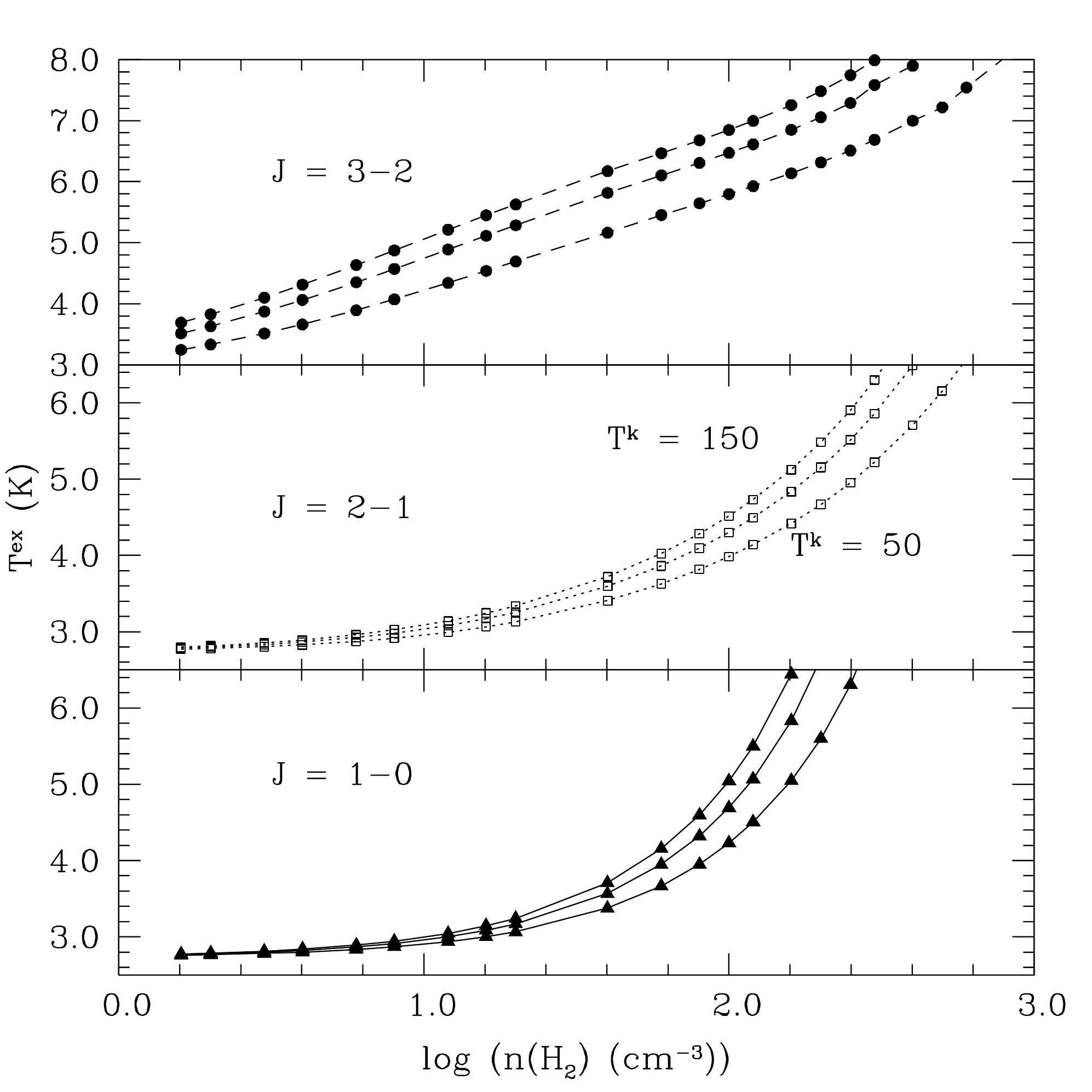}
\caption{\label{tex_all_tk}  Excitation temperatures for the three lowest transitions of CO as function of \hh\ density for three kinetic temperatures representative of this sample of lines of sight through diffuse clouds.   All transitions are assumed to be optically thin.  The curves for each rotational transition are for kinetic temperatures T$^{k}$ = 150 K, 100 K, and 50 K, from left to right.  An ortho--to--para ratio (OPR) equal to 3 has been assumed in all cases.  The collision rate coefficients are those of \citet{yang2010}.
}
\end{center}
\ef
%%%%%%%%%%%%%%%%%%%%%%%%%%%%%%%%%%%%%%%%%

The situation for collisions between hydrogen atoms and CO molecules is less satisfactory.
\h0--CO  collisions were analyzed by \citet{chu1975}, who found cross sections for \h0--CO collisions comparable to those for \hh--CO collisions for small $\Delta J$ which are numerically  the largest.
\citet{green1976} carried out calculations for \h0\ colliding with CO, with quite different results.  
First,  the magnitude of the collision rate coefficients are an order of magnitude smaller (1--2 $\times$10$^{-11}$ cm$^3 $s$^{-1}$ compared to 10$^{-10}$ cm$^3$ s$^{-1}$).
Second,  the \citet{green1976} rate coefficients are very small for $\vert \Delta$J$\vert$ $\ge$ 3 compared to those for $\vert \Delta$J$\vert$ = 1 or 2.  
A more recent calculation \citep{balakrishnan2002} suggested cross sections for \h0--CO collisions a factor $\simeq$ 25 larger in total magnitude (compared to \citet{green1976}), and with large--$\Delta J$ transitions enhanced by up to 1000.  
However, a subsequent reconsideration by \cite{shepler2007} of the interaction potential employed suggested that the results of \citet{balakrishnan2002} are erroneous\footnote {We are indebted to Dr. Balakrishnan for a private communication on this subject advising use of the \citet{green1976} calculations for \h0--CO collisions.}.
This is confirmed by a very recent paper by \citet{yang2013}, which finds results for \h0--CO collisions very similar to those of \citet{green1976}.

In Figure \ref{collision_comparison} we compare the excitation temperatures of the three lowest CO rotational transitions produced by collisions with \h0\ and \hh\ (based on the \cite{green1976} rate coefficients for collisions with \h0\ and the Yang et al. (2010) rate coefficients for collisions with \hh).  
The form of the excitation temperature curves are the same for both types of colliders, but shifted by a factor of $\simeq$ 5 for the two lowest transitions and a factor $\simeq$ 25 for the $J$ = 3-2 transition.
The relative ordering of the $T^{ex}$ is highly sensitive to the density of colliding particles.  
One contributor to this is the fact that the $J$ = 1-0 transition is heading towards a population inversion with equal upper and lower level populations and resulting infinite excitation temperature.  
%For very low densities, the value of $T^{ex}_{J~J-1}$ increases monotonically with $J$, as discussed in the Appendix.
From this we see that for \hh\ fraction $f$(\hh) $\geq$ 0.3, collisions with \hh\ will be dominant, and for $f$(\hh) $\geq$ 0.5, the \h0\ is unimportant for collisional excitation.

Based on the modeling of diffuse molecular clouds in terms of the distribution of \hh\ and \h0\ as a function of optical depth, the CO chemistry, and the relative magnitudes of the collision rate coefficients for collisions with \hh\ and \h0\, as well as the lack of convincing evidence for any non--collisional excitation mechanism, it appears very likely that collisions with \hh\ molecules are the dominant source of collisional excitation of CO molecules.
Consequently, from the observed excitation temperatures we should be able to derive the \hh\ density in diffuse molecular clouds.  
Any result is, of course, subject to the caveat of being an average of the regions with CO along the line of the sight, including multiple clouds, if present. 

%%%%%%%%%%%      FIGURE 4   %%%%%%%%%%%%%%%%%%%%%%%%%%%%%%%
\bf
\begin{center}
\includegraphics[width = 0.8\textwidth]{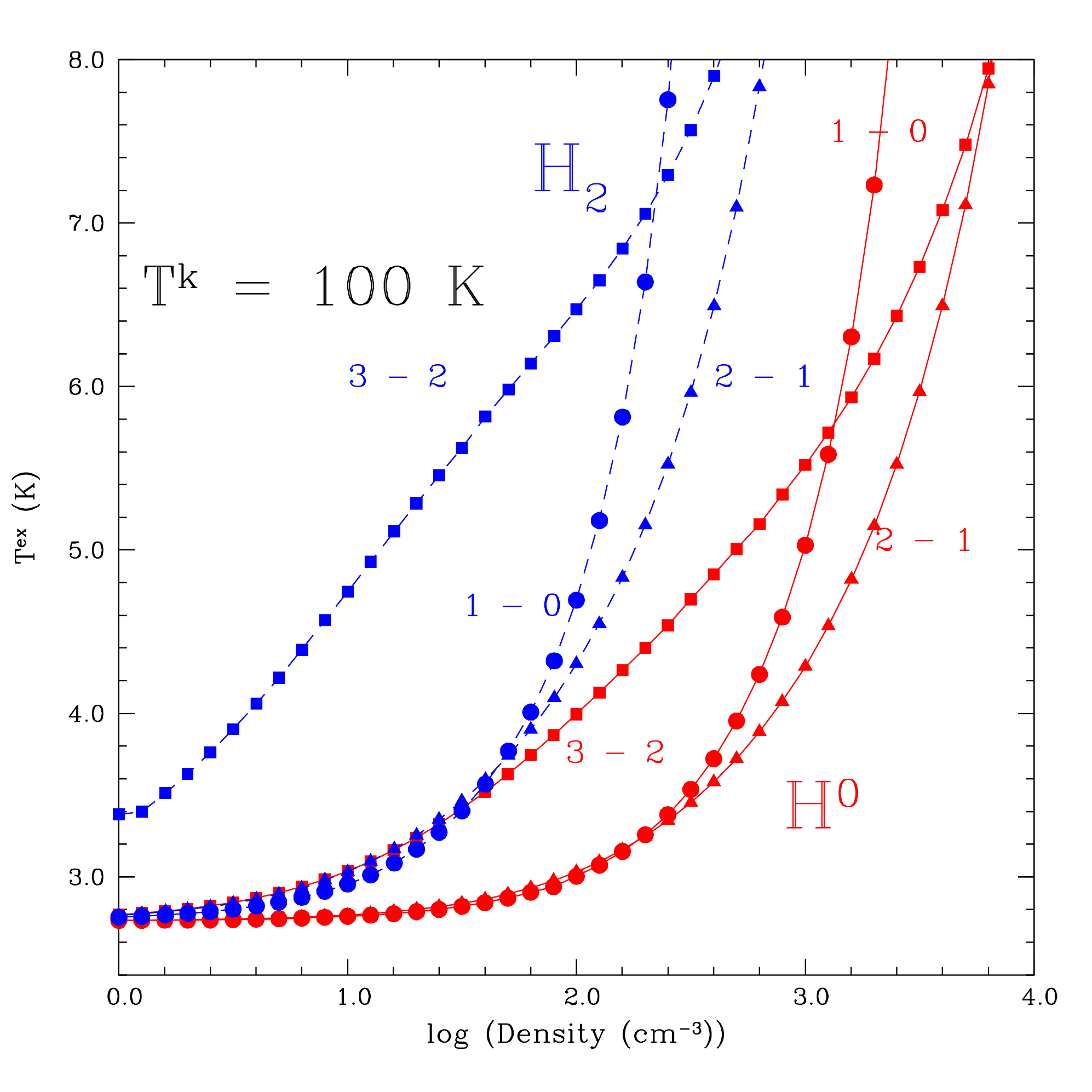}
\caption{\label{collision_comparison} Excitation temperature of the three lowest rotational transitions of CO as a function of colliding particle density.  The kinetic temperature is 100 K and all CO transitions are optically thin.  The J = 3--2 transition is denoted by the squares, the J = 2--1 transition by triangles, and the J = 1--0 transition by circles.  The two sets of transitions correspond to collisions with \hh\ (broken lines) and with \h0\ (solid lines).   A factor $\simeq$ 10  higher $n$(\h0) than $n$(\hh) is required to produce the same value of $T^{ex}_{10}$ or $T^{ex}_{21}$ while a factor $\simeq$ 25 higher \h0\ than \hh\ density is required to produce equal values of $T^{ex}_{32}$.  
}
\end{center}
\ef
%%%%%%%%%%%%%%%%%%%%%%%%%%%%%%%%%%%%%%%%%%%
 \section{Density Determination}
 \label{densities}
 We divide the analysis into two parts.  
 The first is for the sources for which there is only data for the lowest transition.  
 For these sources with only a single value of the excitation temperature to fit, we can determine an upper limit to the density or a range of densities, depending on the value of $T^{ex}_{10}$.
 The second part is for sources with multiple transitions, for which we must also consider the consistency between the results from the different transitions observed.
For the \citet{sheffer2008} data, we consider the uncertainty in the value of $T^{ex}$ to be that given in Section \ref{uncertainties}, namely $\sigma_{T^{ex}}/T^{ex}$ = 0.2, while for the other data we use the uncertainties provided.

\subsection{Optical Depth}
The optical depth of the J = 1--0 CO transition can be written
\be
\tau (1,0) = 3.0\times10^{-4}(\frac{N_{12}}{\delta v_{kms}})(e^{5.53/T^{ex}_{10}} - 1)f_{J=1}\lc
\ee
where $N_{12}$ is the column density of CO in units of 10$^{12}$ cm$^{-2}$, $\delta v_{kms}$ is the FWHM line width in kms$^{-1}$, and $f_{J=1}$ is the fraction of the molecules in the upper ($J$ = 1) state.
Under the subthermal conditions encountered here, it is not correct to assume that all transitions have the same excitation temperature or to adopt the usual expression for the partition function, $Q$ = $KT/hB_0$.
The higher--$J$ transitions have lower optical depths for the densities in the range of interest for these diffuse molecular clouds.
The excitation--dependent terms in the above equation vary among the clouds studied here, but their product is not far from unity.  
While the lines are not spectrally resolved, high--resolution ground-based observations of species such as CH and CH$^+$ as templates suggest FWHM line widths $\simeq$ 3 \kms.
Thus, $\tau(1,0)~\simeq~10^{-4}N_{12}$.
Even if the entire line of sight CO column density is incorporated into a single cloud, the optical depth for almost all clouds included here is considerably less than unity, with the highest column density cloud (having $N_{12}$ = 10$^4$) just reaching this limit.
If the column density is divided among several clouds that each subtends only a small solid angle as seen by 
the others, the radiative trapping will be reduced accordingly.
We thus thus do not consider trapping to be a significant contributor to the CO excitation for the clouds considered here, although this will not be the case, for example, for translucent clouds with larger CO column densities.
 
 \subsection{Kinetic Temperatures}
 \label{tkin}
 The kinetic temperature shows considerable variation among diffuse molecular clouds.
 \cite{savage1977} employed UV observations of \hh\ in the J = 0 and J = 1 rotational levels, and with the assumption that the relative population of these ground rotational levels of para-- and ortho--\hh\ reflects the kinetic temperature, found that clouds with $N$(H$_2$) greater than 10$^{18}$ cm$^{-2}$ have kinetic temperatures between 45 and 128 K, with an average values for 61 stars of 77$\pm$17 (rms) K. 
 \cite{rachford2002} used a similar technique, finding a slightly lower mean value, with $<T^k>$ = 68 K, and a variance of 15 K, although there were three lines of sight having $T^k$ $>$ 94 K.
 \citet{sheffer2008}, again using the same technique, find the average value of the excitation temperature of $J$ =1 relative to $J$ = 0, $<T_{01}($H$_2)>$ = 77$\pm$ 17 K for 56 lines of sight.
 This should be a good measure of the kinetic temperature.
 The range of $T^k$ determined by HI absorption and emission studies \citep{heiles2003} of the Cold Neutral Medium extends to somewhat lower temperatures, but the column density--weighted peak kinetic temperature is 70 K.
 The range 50 K $\leq$ $T^k$ $\leq$ 100 K thus largely covers the measured range of kinetic temperatures determined for  the diffuse molecular clouds considered here.
 
 \subsection{Sources with J = 1--0 Observations Only}
 
Of the 76 sources, 44 are in this category.  
The results are given in Table \ref{dens_1-0_only}, for which we adopt $T^k$ = 100 K.
For the sources for which $T^{ex}_{10}$ together with the statistical uncertainties define a range of allowed densities, we give the minimum and maximum \hh\ densities, $n_{min}$ and $n_{max}$.
 Given the statistical uncertainty in the excitation temperature, and the $T^{ex}_{10}$ {\it vs} $n$(\hh) curve seen in the lower panel of Figure \ref{tex_all_tk}, we consider that we have only an upper limit on the density of a source having $T^{ex}_{10}$ $\leq$ 3.5 K.  
 We denote this maximum density $n_{max}$, and there is no entry for the minimum density $n_{min}$.

As is immediately seen in  Figure \ref{tex_all_tk}, the dependence of the excitation temperature on the kinetic temperature for $n$(H$_2$) $\leq$ 100 \cc\ is much smaller for the $J$ = 1--0 transition than for the higher transitions.  
For most of the density range of interest, the change in log($n$(H$_2$)) required to achieve a particular $T^{ex}$ is no more than 0.1 dex for kinetic temperature changing from 100 K to 50 K, and less than that for the kinetic temperature changing from 100 K to 150 K. 
The \hh\ density required to achieve a given excitation temperature increases as the kinetic temperature decreases due to the reduced excitation rates; the $J$ = 3 level is 33 K above the ground state. 
We adopt a kinetic temperature of 100 K for analysis of the J = 1 -- 0 only clouds.
The modest sensitivity to kinetic temperature indicates that our lack of knowledge of the kinetic temperature in a given cloud or the likely variation in the kinetic temperature throughout a single cloud will not produce a significant error in the derived value of the \hh\ density compared to that resulting from the uncertainty in the excitation temperature arising from the imprecisely known column densities.

 Of the 44 sources with only $T^{ex}_{10}$ data, 30  have only upper limits on $n$(\hh) and 14 have both lower and upper limits.  
 The values of $n_{max}$ for the former are relatively modest, all below 200 \cc, with the average value of $log(n_{max})$ equal to 1.57, corresponding to $<n_{max}>$ $\simeq$ 37 \cc.  
 This category includes, but is not restricted to, clouds having the lowest \hh\ column densities.  
 For each source the logarithm of the midpoint density, $log(n_{mid})$ is the average of the logarithms of $n_{max}$ and $n_{min}$.  
 For the 14 sources with upper and lower limits, the average value of $n_{min}$ is 22 \cc, and of $n_{max}$ is 105 \cc.  
 The average of the midpoint values of $log (n(H_2))$ is 1.69 corresponding to $<n_{mid}>$ = 49 \cc.
 These sources thus represent a population of diffuse clouds having relatively low densities.
 Thermal balance calculations indicate that these low density diffuse clouds will have relative high kinetic temperatures, thus justifying our adoption of 100 K for the nominal value of $T^k$.

 \subsection{Sources with Observations of Two or Three Transitions} 
 
 Our sample includes 18 sources with two, and 14 sources with data for three transitions.  
 The results for these 32 sources are given in Table \ref{multiple}.   
 For each source we give the minimum and maximum density for each transition as discussed above, for kinetic temperature equal to 100 K.  
A dash indicates that there was no excitation temperature for that transition.
In the last two columns we give the range of densities that satisfies all of the data available, if such a consistent range exists.
Sources for which there is an upper limit only for a given transition have no entry in the appropriate $n_{min}$ column.

For sources with data for more than one transition, there is the possibility of no density simultaneously yielding the different excitation temperatures even when the errors are included.
For 18 sources,  we find a range of densities consistent with all transitions observed.
For 7 sources for which there was no formal consistent solution, an additional 0.1 dex in density allows a consistent density or density range to be found. 
These combined densities are indicated by an (s) by the derived density or density range.
The absence of an entry in both of the final two columns indicates there was no density consistent with the excitation temperatures for that source; there are 7 sources in this category.
 
 We have 18 sources with data on $T^{ex}_{10}$ and $T^{ex}_{21}$. 
 Of these, 11 have a density range or upper limit consistent with the measurements of both transitions, while 3 additional sources are in this category if the additional 0.1 dex uncertainty is allowed.  
 For the 12 sources with consistent density ranges, we find $<n_{min}>$ = 22 \cc, and $<n_{max}>$ = 70 \cc.
 The average value of the midpoint densities is $<log(n_{mid})>$ = 1.63 corresponding to $<n_{mid}>$ = 43 \cc.
 This is slightly lower than, but certainly consistent with, the value obtained for the sources for which we have  $T^{ex}_{10}$ data only. 
 This suggests that the two lowest CO transitions are not probing very different regions within or among diffuse molecular clouds along the line of sight.

We have 14 sources with data on excitation temperatures for 3 transitions.  
Since the three different excitation temperatures are differently sensitive to density, these sources are the most demanding in terms of defining a single characteristic density responsible for the entirety of the emission.
Of these 14 sources, 7 have \hh\ density ranges consistent with all three transitions, with 4 additional sources included if we allow the additional 0.1 dex in density added to range for each transition.
For the 11 sources with consistent density range, we find $<n_{min}>$ = 75 \cc, $<n_{max}>$ = 118 \cc, and $<n_{mid}>$ = 94 \cc.
These values are somewhat higher than for the two previous categories, which suggests that inclusion of the $J$ = 3 -- 2 transition does tend to select out clouds or regions within clouds having somewhat higher densities.  
Given the uncertainties, the values of $<n_{mid}>$ of 49, 42, and 94 \cc\ can be taken together to define and average density $<n_{mid}>$ = 60 \cc\ for the 36 sources with \hh\ density ranges, again assuming a kinetic temperature $T^k$ = 100 K.

Of the 32 sources with multiple transition data, we obtain a consistent density ranges for 9 (50\%) of those with the two lowest transitions, and 7 (50\%) of those with three transitions.  
If we include the stretch sources, these numbers go up to 12 (67\%) and 10 (71\%).  
Thus, the inclusion of sources with 3 as compared to 2 transitions leaves the fraction of sources for which a consistent density range can be found essentially unchanged.
Of the 7 sources with no consistent density solution, 5 can be characterized as having the $J$ = 1-0 transition implying too--low density (compared to $J$ = 2-1 (4 sources) or to both $J$ = 2-1 and 3-2 (1 source)).  
The 2 remaining sources are characterized by having $J$ = 2-1 transition implying a density range higher than that indicated by the $J$ = 1-0 and 3-2 transitions.

While the $J$ = 3--2 transition data are suggestive of higher densities, it is not obvious that a density gradient or multiple density components affect level populations in a way that prevents  obtaining a single density solution.  
In fact, the simple combination of two densities generally produces a solution that is simply an intermediate value.
This is illustrated in Figure \ref{2comp}, in which we have combined two different clouds having densities of 10 \cc\ and 100 \cc\  with the low density component (cloud 1) having a fraction between 0 and 1 of the total CO column density.  
We assume that both clouds have the same kinetic temperature.
Since all lines are optically thin, it is straightforward to calculate the excitation temperatures that would be derived from the relative column densities.
The result is that the variation in the three excitation temperatures produced by varying the relative amount of high and low density cloud material mimics quite closely the variation in the excitation temperatures produced by a single cloud having density between that of the lower density and the higher density cloud.  
%%%%%%%%%%%      FIGURE 5  %%%%%%%%%%%%%%%%%%%%%%%%%%%%%%%
\bf
\begin{center}
\includegraphics[width = 0.7\textwidth]{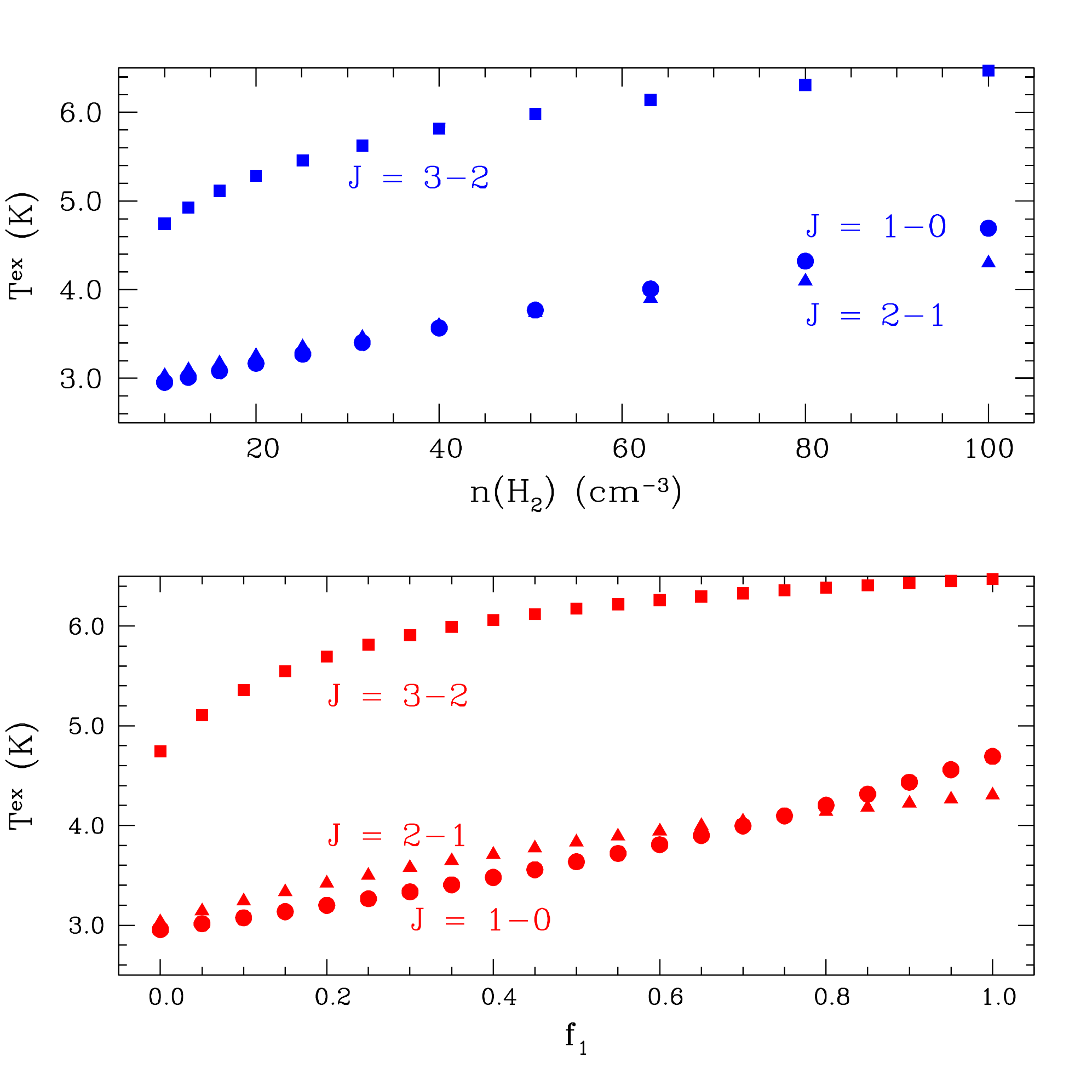}
\caption{\label{2comp} Comparison between varying the density within a single region and combining two regions having different densities.  The kinetic temperature in all cases is 100 K and the background temperature is 2.7 K.  Upper panel:   excitation temperatures for three lowest CO transitions as function of \hh\ density in a single region producing optically thin emission with 10 \cc\ $\leq$ $n$(H$_2$) $\leq$ 100 \cc.   Lower panel:  excitation temperatures resulting from combination of two regions.  Region 1 has $n$(\hh) = 10 \cc\ and region 2 has $n$(\hh) = 100 \cc.  The fraction of the total column density in region 1 is denoted $f_1$.  The three excitation temperatures show almost exactly the same variation in both cases, suggesting that a combination of 2 clouds with high and low densities is not distinguishable from a single cloud having an intermediate density.
}
\end{center}
\ef
%%%%%%%%%%%%%%%%%%%%%%%%%%%%%%%%%%%%%%%%%%%

\subsection{Average Density and Thermal Pressure}

The values for the \hh\ density of each source have been found in terms of maximum and minimum values of log(n(\hh)) that are consistent with the data including errors predominantly due to statistical uncertainties.  
The sources having 2 or 3 values of kinetic temperature are likely to be the most valuable for assessing the effect of kinetic temperature changes since the different transitions have upper levels significantly higher than for the $J$ = 1--0 transition, although there are relatively fewer of the multiple--transition sources.  
Figure \ref{tex_all_tk} shows that while $T^{ex}_{10}$ is relatively insensitive to $T^k$, the higher transitions show increasing sensitivity, as expected for the larger level separation (equivalent to 33 K for the $J$=3--2 transition).  
The collision rates increase monotonically with kinetic temperature in the range of interest, and thus the density required to obtain a given kinetic temperature is lower for a higher value of $T^k$.  

In Figure \ref{tkin_comparison} we show graphically the range of densities for each of the excitation temperatures in six sources in this category.
As anticipated, the allowed densities are shifted to higher values for the lower kinetic temperature.  
This applies to the individual transitions as well as for the allowed ranges for the combined set of three transitions.
For four of the six sources, the allowed density range for the combined set of transitions is substantial.
However, for HD148937, the combined transition density range is very narrow, only 0.1 dex.
For HD147683 there is nominally no density consistent with all three excitation temperatures, but $log(n($H$_2$)) is within 0.1 dex of the upper limit from the J = 1--0 transition and an equal amount from the lower limit of the $J$ = 2--1 transition for $T^k$ = 100 K, and similarly $log(n($H$_2$)) = 2.5 for $T^k$ = 50 K.  
There is no obvious pattern from changing the kinetic temperature other than the shift to slightly higher densities for 50 K compared to 100 K kinetic temperature.  
It therefore does not seem possible to use the available data to put tighter constraints on the kinetic temperature; measurements of higher--$J$ transitions would be required to do this.

We can find the average value of the midpoint density for several different groupings of our sources, and the results are given in Table \ref{average_densities}. 
We see that $<n_{mid}>$ is essentially the same for the 14 sources for which we have only $T^{ex}_{10}$ and the 12 sources for which we have values for  $T^{ex}_{10}$ and $T^{ex}_{21}$.  
For both categories,  $<n_{mid}>$ $\simeq$ 45 \cc, at a kinetic temperature of 100 K.
The minimum, maximum, and midpoint densities are all greater when we consider 3 rather than 2 excitation temperatures, as discussed above.
If we include the 23 sources with 2 or 3 excitation temperatures the average value of the midpoint density is $<n_{mid}>$ = 68 \cc, compared to 42 \cc\ for two excitation temperatures, and 94 \cc\ for three excitation temperatures, all for $T^k$ = 100 K.

For a lower kinetic temperature of 50 K, we obtain somewhat higher densities.  
For the sources with data on two excitation temperatures, $<n_{min}>$ = 32 \cc, $<n_{mid}>$ = 67 \cc, and $<n_{max}>$ = 143 \cc.  
For the sources with data on three excitation temperatures, $<n_{min}>$ = 104 \cc, $<n_{mid}>$ = 135 \cc, and $<n_{max}>$ = 176 \cc.
All of these results are in line with previous determinations of densities of diffuse clouds.  
There do seem to be clear variations among the sources included in this study, with some sources having n(\hh) only a few tens \cc\ (HD23478 and HD24398), while HD147888 has  a density at least a factor of 10 higher.

The thermal pressure suggested by these results is moderately large.
The anticorrelation between assumed $T^k$ and derived n(\hh) suggests that a thermal pressure derived by taking their product is reasonably robust against errors in the kinetic temperature. 
Using the midpoint densities for the sources with two or  three values of excitation temperature as the largest statistical sample with reasonable sensitivity to kinetic temperature, we find for $T^k$ = 100 K, $p/k$ = 6800 K\cc, and for $T^k$ = 50 K, $p/k$ = 4600 K\cc.  
%
%	Something may be amiss is averaging here.  It should be 5700  not 6700!!!  The range 4600 to 6800 is correct and is also given correctly in the Abstract and in the Discussion & Summary
%
Further taking the average of these two yields a thermal pressure $p/k$ = 6700 K\cc.  
This value is noticeably above the median value determined from UV absorption studies of CI by \citet{jenkins2001}, 2240 K\cc, but within the range of the sources studied similarly by \citet{jenkins2002}, 10$^3$ K\cc\ $\leq$ $p/k$ $\leq$ 10$^4$ K\cc.
A more comprehensive \C1\ study of 89 stars by \citet{jenkins2011} finds a lognormal pressure distribution with $<log(p/k)>$ = 3.58, corresponding to $<p/k>$ = 3800  K\cc.
It is possible that while the densities found here from CO are still quite modest, the regions may be the envelopes of molecular clouds, which are characterized by significantly higher thermal pressure than for diffuse molecular clouds \citep{wolfire2010}.

\subsection{Correlation Between Volume Density and Column Density}
The present data allow us to examine whether there is a correlation between volume density and column density for this sample of diffuse clouds.
We have 14 sources with density ranges from $T^{ex}_{10}$ alone and molecular hydrogen column densities.  
These are plotted with diamond (black) symbols in  Figure \ref{dens_coldens}.
We also have 16 sources from our sample with density ranges determined by excitation temperatures from 2 (9 sources) and 3 (7 sources) transitions; these are plotted with square (red) symbols.

For the $T^{ex}_{10}$ only sources, there is no significant correlation of volume and column density.
The data for the multiple--transition sources suggests a weak correlation, with a linear best fit $n$(\hh) rising from 10 \cc\ to 100 \cc\ as $N$(\hh) increases from 10$^{20}$ \c2\ to 10$^{21}$ \c2. 
It is clear, however, that a linear relationship is not consistent with the data for HD 147888, which has a density $\simeq$ 4 higher than the general trend.
These data suggest that there is at least a component of diffuse molecular clouds for which volume density and column density are correlated.
%
%%%%%%FIGURE 6 %%%%%%%%%%%%%%%%%%%%%%%%%%
\bf
\begin{center}
\includegraphics[width = 0.8\textwidth]{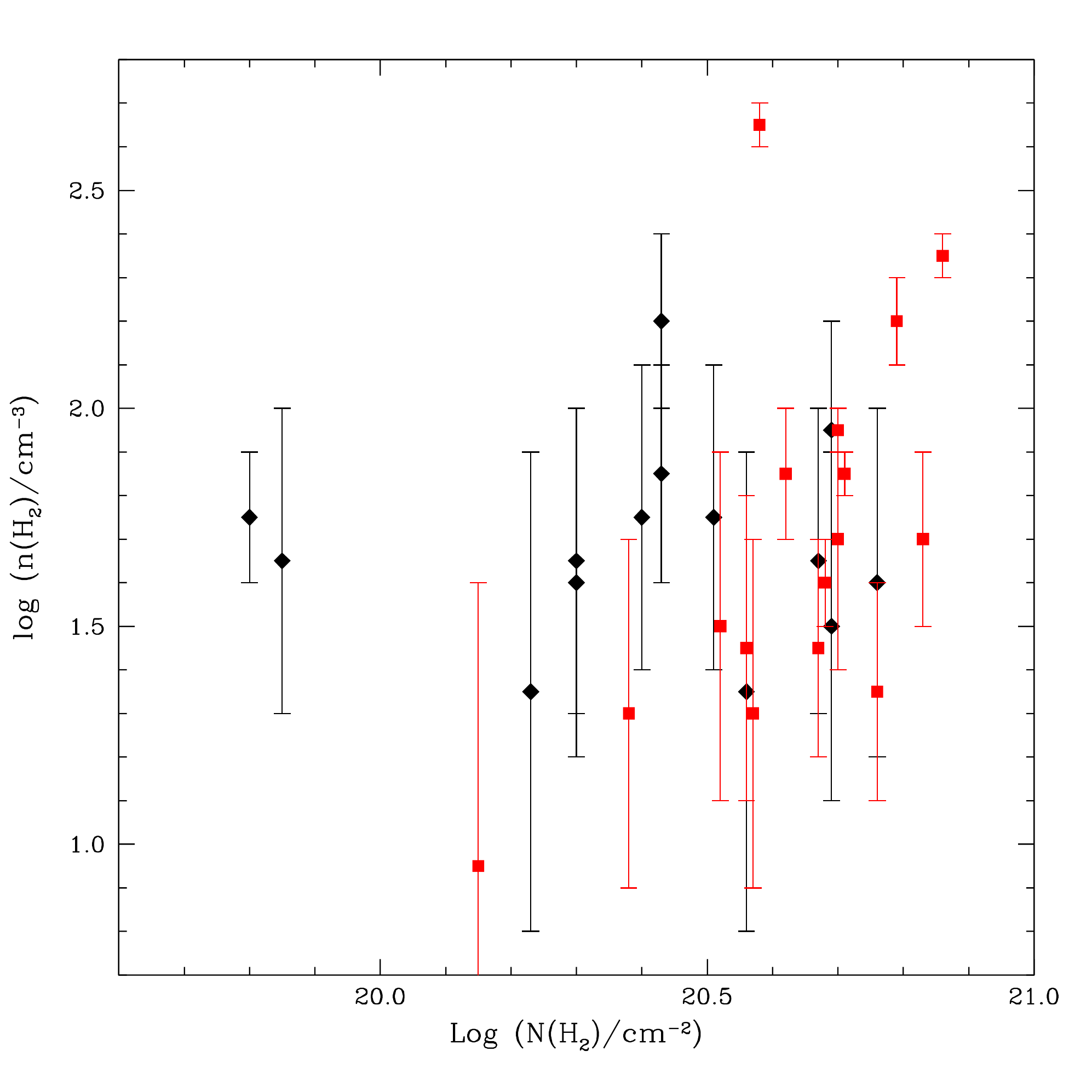}
\caption{\label{dens_coldens}  Volume density of \hh\ determined from CO absorption as a function of \hh\ column density.  The sources included here have upper and lower density limits.  Those from the $J$ = 1--0 transition alone are plotted with diamond (black) symbols and those from sources with 2 or 3 excitation temperatures are plotted with square (red) symbols.  The midpoint density for each source is indicated by the symbol and the upper and lower limits of the density range (Table \ref{multiple}) by the error bars.  The source with the unusually high volume density is HD 147888.  The sources with multiple transition data clearly show $n$(\hh) correlated with $N$(\hh), while those with $J$ = 1--0 only data do not.
}
\end{center}
\ef
%%%%%%%%%%%%%%%%%%%%%%%%%%%%%%%%%%%%%%
%%
%%%%%%%%%%FIGURE 7  %%%%%%%%%%%%%%%%%%%%%%  
\bf
\begin{center}
\includegraphics[width = 0.8\textwidth]{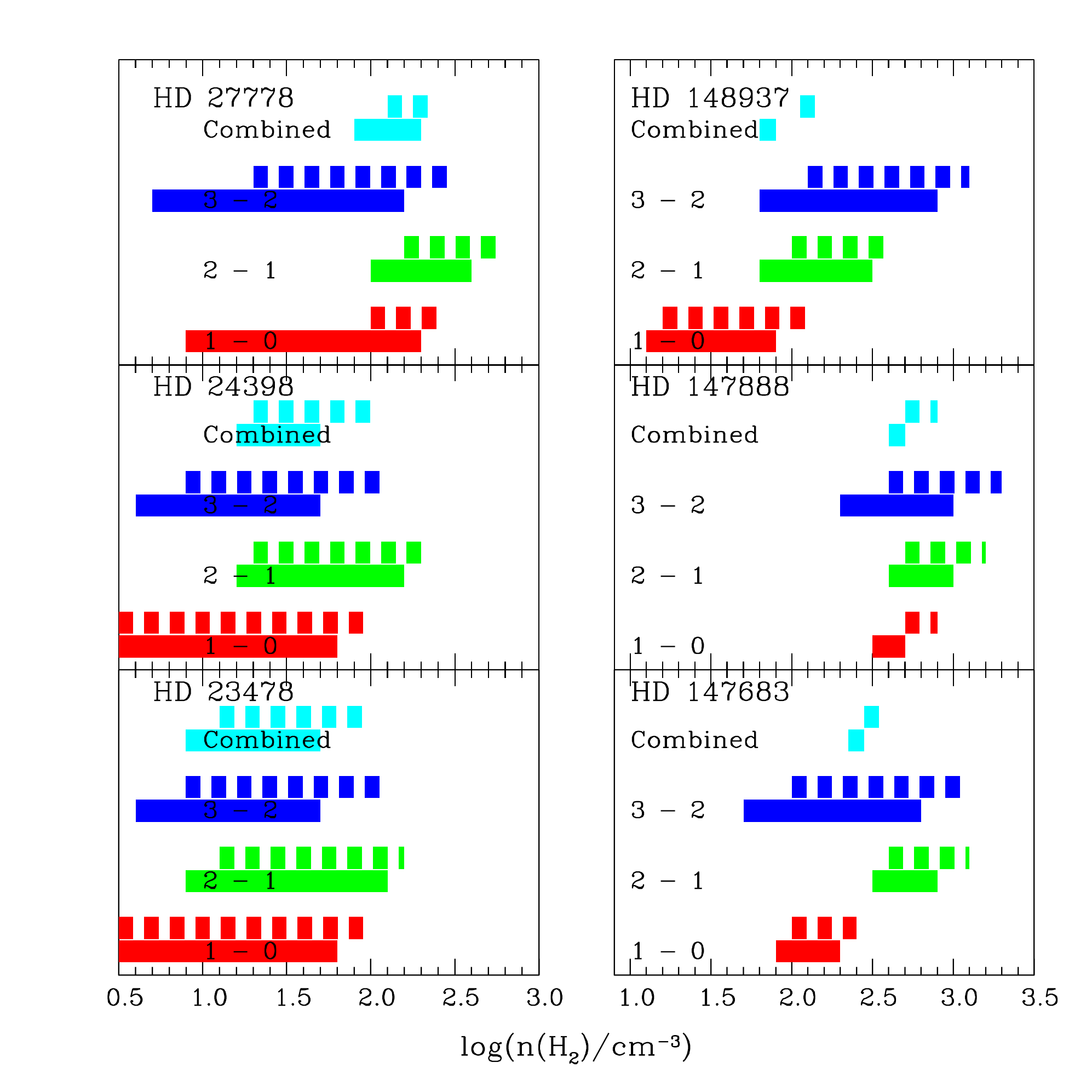}
\caption{\label{tkin_comparison}  Volume density of \hh\ determined from CO absorption towards six of the sources for which 3 excitation temperatures are determined.  For each source, the ranges permitted (given assumed $\pm$20\% uncertainties) for each transition are indicated by the bars (red, green, and blue) in order of increasing transition frequency.  The ranges for $T^k$ = 100 K are indicated by the solid bars, and those for 50 K by the dashed bars plotted just above.}
\end{center}
\ef
%%%%%%%%%%%%%%%%%%%%%%%%%%%%%%%%%%%%%%

\section{Discussion and Summary}
\label{discussion}

We have used the UV CO absorption data of \cite{sheffer2008} and published data on other sources, together with statistical equilibrium calculations, to determine the volume density in diffuse interstellar molecular clouds.
We have a total of 76 sources,  of which 44 have $T^{ex}_{10}$ data only, 18 sources having  $T^{ex}_{10}$ and  $T^{ex}_{21}$ data , and 14 sources having  $T^{ex}_{10}$, $T^{ex}_{21}$, and $T^{ex}_{32}$ data.  
It does not appear likely that non--collisional processes play a major role in excitation of CO in diffuse clouds.
Collisional excitation is expected to result primarily from collisions with \hh\ molecules as the H$^0$ to H$_2$ transition occurs at substantially lower values of column density than does the C$^+$--C$^o$--CO transition.  
Recent calculations confirm that excitation rate coefficients for CO--\hh\ collisions are significant larger than for CO--H$^0$ collisions.
The CO and H$_2$ column densities of the sources indicate that the fractional abundance of CO is several orders of magnitude below its asymptotic value in well--shielded regions, and also that the CO rotational transitions are optically thin, with the sources having the largest CO column densities reaching $\tau$ $\simeq$ 1.
We have assumed a kinetic temperature of 100 K as representative for the more diffuse clouds, but also discuss the effect of $T^k$ a factor of 2 lower, especially for analysis of sources having data on higher --$J$ transitions.
 
For 30 of the sources having only $T^{ex}_{10}$ data, we obtain only upper limit to $n($H$_2$) for which the average value is $<log(n_{max})>$ = 1.57, corresponding to $<n_{max}>$ = 37 \cc.
For the remaining 14 $T^{ex}_{10}$--only sources we find a range of \hh\ densities that is consistent with the value of the excitation temperature and its estimated uncertainty, and thus determine $n_{min}$ as well as $n_{max}$.
For these sources we find $<n_{min}>$ = 22 \cc\ and $<n_{max}>$ = 105 \cc.
Defining $log(n_{mid})$ as the average of $log(n_{max})$ and $log(n_{min})$ for each source, the average midpoint density for these 14 sources is given by $<n_{mid}>$ = 49 \cc.
Of the 18 sources having $T^{ex}_{21}$ and $T^{ex}_{10}$ data, 14  yield a consistent density range or upper limit.
For the 12 sources with density ranges, $<n_{mid}>$ = 42 \cc\ for $T^k$ = 100 K and 67 \cc\ for $T^k$ = 50 K.
Of the 14 sources having $T^{ex}_{10}$, $T^{ex}_{21}$, and $T^{ex}_{32}$ data, 11 yield density ranges that are consistent for all three transitions, yielding $<n_{mid}>$ = 94 \cc\ for $T^k$ = 100 K and 135 \cc\ for $T^k$ = 50 K.
Taking the sources with either two or three values of the excitation temperature, we find $<n_{mid}>$ = 68 \cc\ for $T^k$ = 100 K and 92 \cc\ for $T^k$ = 50 K.

Thus, while there are undoubtedly some selection biases, it appears that this sample of diffuse molecular clouds, having \hh\ column densities between few $\times$10$^{20}$ and $\simeq$ 10$^{21}$ cm$^{-2}$ is reasonably characterized by a density between 50 and 100 \cc. 
The clouds in this sample clearly do not all have the same volume density, with the extreme cases being a factor of a few below and above the range given here.
The anticorrelation between derived density and the assumed kinetic temperature allows plausible determination of the internal thermal pressure of these clouds, which is found to be relatively large with $p/k$ in the range 4600 to 6800 cm$^{-3}$K.
As we are analyzing clouds in which hydrogen is largely molecular, but in which the fractional abundance of CO is so small that this species would be extremely difficult to detect in emission, the present results help characterize the ``CO--Dark Molecular Component'' of the interstellar medium.

%%%%%%%%%%%%%%%%%%%%%%%%%%%%%%%%%%%%%%%%%%%%%%%%%

We thank Drs. N. Balakrishnan and L. Wiesenfeld for very helpful information about collision rate coefficients and potential energy surfaces.
We thank Nicolas Flagey and Jorge Pineda for useful discussions about dealing with the uncertainties in the column densities of CO and the rotational excitation temperatures, and Bill Langer for clarifying a number of points and a careful reading of the manuscript.  The anonymous referee also made significant contributions by pointing out particular aspects of UV studies of diffuse clouds that would otherwise have been missed, and by carefully checking of the data presented here.  This research was carried out at the Jet Propulsion Laboratory, California Institute of Technology, under contract with the National Aeronautics and Space Administration.
%%%%%%%%%%%%%%%%%%%%%%%%%%%%%%%%%%%%%%%%%%%%%%%%%

\section{Appendix}
\label{appendix}

\subsection{INTRODUCTION}
This appendix addresses the issue that the excitation temperature of the levels of simple molecules and atoms
in the limit of very low densities does not asymptotically approach the temperature of the background radiation field.
This applies to rigid rotor molecules such as CO, and also to simple atomic systems such as \C1\ and \O1.

Using the RADEX program \citep{vandertak2007} to analyze CO excitation by collisions with ortho--\hh\ molecules for a kinetic temperature of 100 K, background temperature of 2.7 K, and \hh\ density of 0.01 \cc\ yields the results shown in Table \ref{radex_co}.  
All transitions are optically thin, and since collisional deexcitation rate coefficients are $\simeq$ few $\times$\ 10$^{-11}$ cm$^{3}$s$^{-1}$, all transitions should be highly subthermal, given that the $A$-coefficient for the lowest transition is 7.2$\times$10$^{-8}$ s$^{-1}$.

%\begin{center}

%\end{center}

\normalsize
This result, that the excitation temperatures seem unreasonably large and increase with increasing $J$ (albeit not perfectly monotonically), is found in the output of all statistical equilibrium programs examined.  
It thus does not seem to be an artifact of the calculation, but rather is a property of the solutions of the rate equations in the low density limit.  
While this may seem to be a curiosity, it is important if one has level populations derived from UV absorption, for example, and one wishes to solve for densities that result in highly subthermal excitation.  
This is suggested by the FUSE and HST observations of CO of \citet{sheffer2008} and others that are discussed in this paper.

\subsection{Three Level Model}
\subsubsection{Definitions}

In order to gain some insight into the behavior of the excitation temperatures, we can use a three level model with two transitions to capture the essence of the multilevel CO problem. 
This is a complete representation of the situation for atomic \C1\ and \O1\ fine structure systems and a very good approximation for CO at low densities. 
We denote the levels 1, 2, and 3 (not to be confused with rotational quantum numbers), their energies as $E_1$, $E_2$, and $E_3$, and the downwards spontaneous rates and collision rates as $A_{21}$, $A_{32}$, $C_{21}$, $C_{32}$, and $C_{31}$.  
The energies of the three levels lead to equivalent temperatures for the three transitions $kT_{21} = E_2 - E_1$, $ kT_{32} = E_3 - E_2$, and $kT_{31} = E_3 - E_1$, where $k$ is Boltzmann's constant.
The background radiation field is assumed to be a blackbody at temperature $T^{bg}$ producing energy density $U(T^{bg})$.
The downwards stimulated rates are $B_{21}U$ and $B_{32}U$, where the $B$'s are the stimulated radiative rate coefficients and $U$ is understood to be evaluated at the frequency of the transition in question.
The upwards stimulated rates are related to the downwards rates through the statistical weights $g_1$, $g_2$, and $g_3$ and detailed balance, giving $g_1B_{12}U = g_2B_{21}U$ and $g_2B_{23}U = g_3B_{32}U$.

The collision rates are the product of the collision rate coefficients and the colliding partner density.
Thus for collisions with \hh,  
\be
\label{collrates}
C_{ij} = R_{ij}n(H_2)
\ee
The upwards rates are related to the downwards rates through detailed balance and the kinetic temperature $T_k$ through $g_1C_{12} = g_2C_{21}$exp$(-T^*_{21}/T^k)$, $g_1C_{13} = g_3C_{31}$exp$(-T^*_{31}/T^k)$, and $g_2C_{23} = g_3C_{32}$exp$(-T^*_{32}/T^k)$.

The level population per statistical weight defines the excitation temperature through equation \ref{nunl}.
With these definitions, the rate equations for the level populations $n_1$, $n_2$, and $n_3$ lead to the following equations for the ratios of the column densities of adjacent levels (connected by radiative transitions):

\be
\label{3lev_n21}
\frac{N_2}{N_1} = \frac{(A_{32}+B_{32}U+C_{32}+C_{31})(B_{12}U+C_{12}+C_{13})-C_{13}C_{31}}{(A_{32}+B_{32}U+C_{32}+C_{31})(A_{21}+B_{21}U+C_{21})+(C_{23}+ B_{23}U)C_{31}} \lc
\ee
and
\be
\label{3lev_n32}
\frac{N_3}{N_2} = \frac{(C_{23}+B_{23}U)(B_{12}U+C_{12}+C_{13})+(A_{21}+B_{21}U+C_{21})C_{13}}{(A_{32}+B_{32}U+C_{32}+C_{31})(B_{12}U+C_{12}+C_{13})-C_{13}C_{31}} \lp
\ee

\subsubsection{High Density Limit}

In this limit with $C$ $\gg$ $A$, $BU$, we find that equations \ref{3lev_n21} and \ref{3lev_n32} yield $N_2/N_1 = (g_2/g_1)e^{(-T^*_{21}/T^{k})}$ and $N_3/N_2 = (g_3/g_2) e^{(-T^{*}_{32} /T^{k})}$, respectively.  
This is exactly as expected in the thermalized limit when collisions dominate.

\subsubsection{Zero Collision Rate Limit}

In this limit
\be
\label{3lev_n21_nocoll}
\frac{N_2}{N_1} = \frac{B_{12}U}{A_{21}+ B_{21}U} \lc
\ee
and
\be
\label{3lev_n32_nocoll}
\frac{N_3}{N_2} = \frac{B_{23}U}{A_{32}+B_{32}U} \lp
\ee
which yield $T^{ex}$ = $T^{bg}$ for both transitions.

\subsubsection{Low Density Limit with No Background Radiation}

With the above expressions we can examine the low density limit, together with the effect of varying the background radiation temperature.
We first consider the no--background limit ($T^{bg}$ = 0).
In this case, dropping collisional terms where they compete directly with a spontaneous rate, we find
\be
\label{3lev_n21_tbg0_lowden}
\frac{N_2}{N_1} = \frac{C_{12}+C_{13}}{A_{21}} \lc
\ee
and
\be
\label{3lev_n32_tbg0_lowden}
\frac{N_3}{N_2} = \frac{A_{21}C_{13}+C_{23}(C_{12}+C_{13})}{A_{32}(C_{12}+C_{13})} \lp
\ee
%Strictly speaking, the second term in the numerator of the expression for $N_3/N_2$ should be dropped, but we have retained it temporarily to indicate the critical role played by nonzero value for collisions connecting levels 1 and 3.  
If the $\Delta J$ = 2 collision rate coefficients  are zero, equation \ref{3lev_n32_tbg0_lowden} reduces to
\be
\label{n32_dipole}
\frac{N_3}{N_2} = \frac{C_{23}}{A_{32}} \lp
\ee
In this case of purely ``dipole--like'' collisions, we also find
\be
\label{n21_dipole}
\frac{N_2}{N_1} = \frac{C_{12}}{A_{21}} \lc
\ee
We thus see that the excitation temperature of each transition approaches zero as the collision rate approaches zero.  

If collisions between levels 1 and 3 are allowed, then in the limit of very low collision rate, we find
\be
\label{n32_lowcoll_nobg}
\frac{N_3}{N_2} = \frac{A_{21}}{A_{32}}\frac{C_{13}}{C_{12}+C_{13}} \lp
\ee
This obviously has an entirely different behavior than that of the lower transition given by equation \ref{3lev_n21_tbg0_lowden}.  
The excitation temperature of the upper transition approaches an asymptotic limit since the first fraction is a constant determined by the molecular radiative rates, and the second fraction is a constant determined by the relative collision rates.  
For the latter, in the limit of zero background, the excitation temperature for the upper transition is \textbf{independent of the density}, and is given by
\be
\label{tex32_lowden}
T^{ex}_{32} = {T^*_{32}}/{ln[\frac{g_3 A_{32}}{g_2 A_{21}} (\frac{R_{12}+R_{13}}{R_{13}})] }\lp
\ee

In the low density limit (with no background), the excitation temperature for the lower transition is
\be
\label{tex21_lowden}
T^{ex}_{21} = {T^*_{21}}/{ln[\frac{g_2}{g_1} \frac{A_{21}}{C_{12}+C_{13}}]} \lc
\ee
which \textbf{does} depend on the collision partner density through the proportionality of the collision rates and the density (equation \ref{collrates}).

To restate the obvious, the excitation temperature of the upper (level 3 -- level 2) transition does not approach zero even if the collision rate is arbitrarily small.  
This is because there is not a simple competition between collisional and radiative processes.  
This is in contrast with the lower (level 2 -- level 1) transition, for which the excitation temperature does approach zero for low collision rate.  
This reflects the fact that in the limit of very infrequent collisions, level 2 is populated exclusively by collisions from level 1 (which has most of the population) and depopulated by radiative decay back to level 1.  

It is thus evident that the excitation temperature of a particular transition in a multilevel system can behave in the apparently counterintuitive way of having $T^{ex}$ not approach zero as the collision rate approaches this value.  We next give some examples of three-level systems, and will extend the discussion to systems with more than 3 levels in Section \ref{multilevel}

\subsubsection{Examples of Different Systems and Comparisons with Numerical Calculations}
\label{comparison}

\subsubsection{CO}

We consider the lowest three rotational levels of CO to illustrate the preceding analytic results.  
The rate coefficients for collisions with para--\hh\ from \cite{yang2010} at a kinetic temperature of 100 K are $R_{12} = 9.7\times10^{-11}$ cm$^3$s$^{-1}$ and $R_{13} = 1.4\times10^{-10}$ cm$^3$s$^{-1}$.
Since the collision rates and rate coefficients are proportional (equation \ref{collrates}), this gives $(R_{12}+R_{13})/R_{13}$ = 1.7, which yields (for no background radiation) $T^{ex}_{32} = 3.35$ K.  
%This agrees quite closely with the value obtained from RADEX in the limit of very small background temperature.  
The collisional excitation rates for ortho--\hh\ -- CO collisions from \citet{flower2001} and \citet{wernli2006} as extrapolated in the LAMDA database (home.strw.leidenuniv.nl/~moldata/) at a kinetic temperature of 100 K are $R_{12} = 2.65\times10^{-10}$ cm$^3$s$^{-1}$ and $R_{13} = 2.63\times10^{-10}$ cm$^3$s$^{-1}$.  
This gives $(R_{12}+R_{13})/R_{13}$ = 2.1 and an excitation temperature (for no background radiation) $T^{ex}_{32} = 3.2$ K.  
The value from the full multilevel RADEX calculation is 3.6 K.  
The difference is due to the effect of the higher levels, discussed in Section \ref{multilevel}.

 $T^{ex}_{32}$ is essentially constant for \hh\ densities up to 100 \cc, at which point it begins to rise due to the collision rate becoming comparable to the spontaneous decay rate.
The excitation temperature of the lower transition, as expected, varies continuously as a function of the \hh\ density.
This behavior is shown in Figure \ref{tex_nh2}.  
The excitation temperature of the lower transition is rising sharply as $n$(\hh) approaches 100 \cc, because with the relatively large rate for $\Delta J$ = 2 collisions, we can have a situation in which level 3 ($J$ = 2) is populated by collisions from level 1 ($J$ = 0).  
The radiative decays to level 2 ($J$ = 1) add to the direct collisional population of that level and result in level 2 ($J$ = 1) becoming overpopulated relative to level 1($J$ = 0) as seen in \citet{goldsmith1972}. 
As the \hh\ density increases, the negative excitation temperatures characteristic of the population inversion are preceded by very high positive values of the excitation temperature of the lowest transition, $T^{ex}_{21}$.  

The behavior of  $T^{ex}_{32}$ is largely independent of the choice of collision partner or which calculation of the collision rate coefficients is adopted.  
The \cite{green1976} rate coefficients for CO--\hh\ collisions give $(R_{12}+R_{13})/R_{13}$ = 1.61 and $T^{ex}_{32}$ = 3.4 K, almost identical to the results from \cite{yang2010}, although the values for the individual coefficients are slightly larger.
\cite{green1976} also give the results for collisions with H and He atoms, which give $T^{ex}_{32}$ = 3.3 K and 3.47 K, respectively, almost identical to the values produced by collisions with \hh\ molecules.
%%
%%%%%%%%%%%%FIGURE 8  %%%%%%%%%%%%%%%%%
\begin{figure}[h]
\begin{center}
\includegraphics[width = 0.8\textwidth]{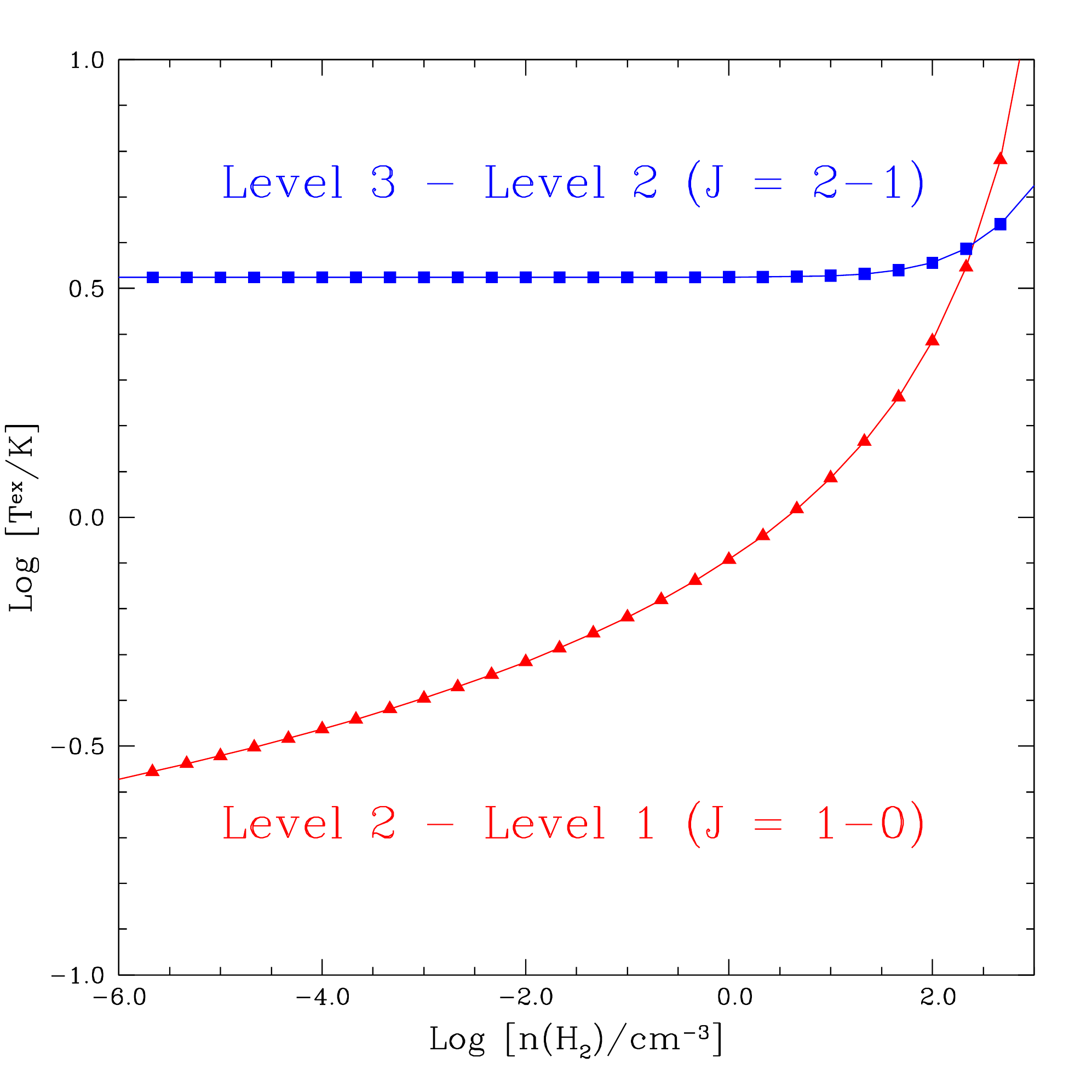}
\caption{\label{tex_nh2} Excitation temperature of two lowest rotational transitions of CO as a function of \hh\ density.  There is no background radiation field, and the kinetic temperature is 100 K.  
}
\end{center}
\end{figure}
%%%%%%%%%%%%%%%%%%%%%%%%%%%%%%%%%%%

\subsubsection{\C1}

The three fine structure levels of \C1\ make this system a highly appropriate test of this behavior for low collision rates. 
The ground state (level 1) is $^3P_0$, the first excited state (level 2 at $E/k$ = 23.65 K above the grounds state) is $^3P_1$, and the second excited state (level 3 at 62.51 K above the ground state) is $^3P_2$.
Adopting the deexcitation rate coefficients of \cite{schroeder1991} for collisions with ortho--\hh, we find for a kinetic temperature of 100 K that $R_{12} = 1.68\times10^{-10}$ cm$^3$s$^{-1}$ and $R_{13} = 1.85\times 10^{-10}$ cm$^{3}$s$^{-1}$.  
With $A_{32} = 2.65\times10^{-7}$ s$^{-1}$ and $A_{21} = 7.9\times10^{-8}$ s$^{-1}$, we find from equation \ref{tex32_lowden} that $T^{ex}_{32}$ = 16.7 K.  
The excitation temperature of the lower transition from equation \ref{tex21_lowden} is 3.65 K for a hydrogen density of 1 \cc\ and no background radiation.  
The value for the upper transition agrees within a few tenths K with that from RADEX, and that of the lower transition agrees within 0.1 K.

\subsubsection{Effect of Background Radiation}
\label{background}

%Retaining the stimulated emission and absorption terms in equations \ref{3lev_n21} and \ref{3lev_n32} makes it more difficult to see clearly what is happening when the collision rate is comparable to the radiative rates.

The ratio of the downwards stimulated emission rate due to the background radiation field to the spontaneous decay rate is given by
\be
\label{stimspon}
\frac{B_{ul}U}{A_{ul}} = \frac{1}{exp(T^*/T^{bg}) - 1} \lc
\ee
where $T^{bg}$ is temperature of the background radiation field, which we assume to be a blackbody.
Let us consider the situation in which $C_{ul} \ll A_{ul}$ with no background radiation field ($T^{bg}$ = 0).
If we consider increasing the background temperature, we will reach a point at which $B_{ul}U = C_{ul}$.  
This occurs when
\be
\label{bg_eq_A}
T^{bg'} = \frac{T^*}{ln(1 + A_{ul}/C_{ul})} \lp
\ee
The required background temperature thus depends on how much smaller the collision rate is than the spontaneous rate.
For this value of background temperature, we should expect the excitation temperature to approach the background temperature since spontaneous and stimulated rates are both comparable to or greater than the collision rate.
The results for the 3 level model for CO are shown in Figure \ref{tex_tbg}, for a kinetic temperature of 100 K and a \hh\ density of 1 \cc.  
At this density, $A_{21}/C_{21}$ = 2.1$\times$10$^3$ and $A_{32}/C_{32}$ = 1.1$\times$10$^4$.
%%
%%%%%%%%%%%%%%FIGURE 9  %%%%%%%%%%%%%%%%%%%%%%%%%%
\begin{figure}
\begin{center}
\includegraphics[width = 0.8\textwidth]{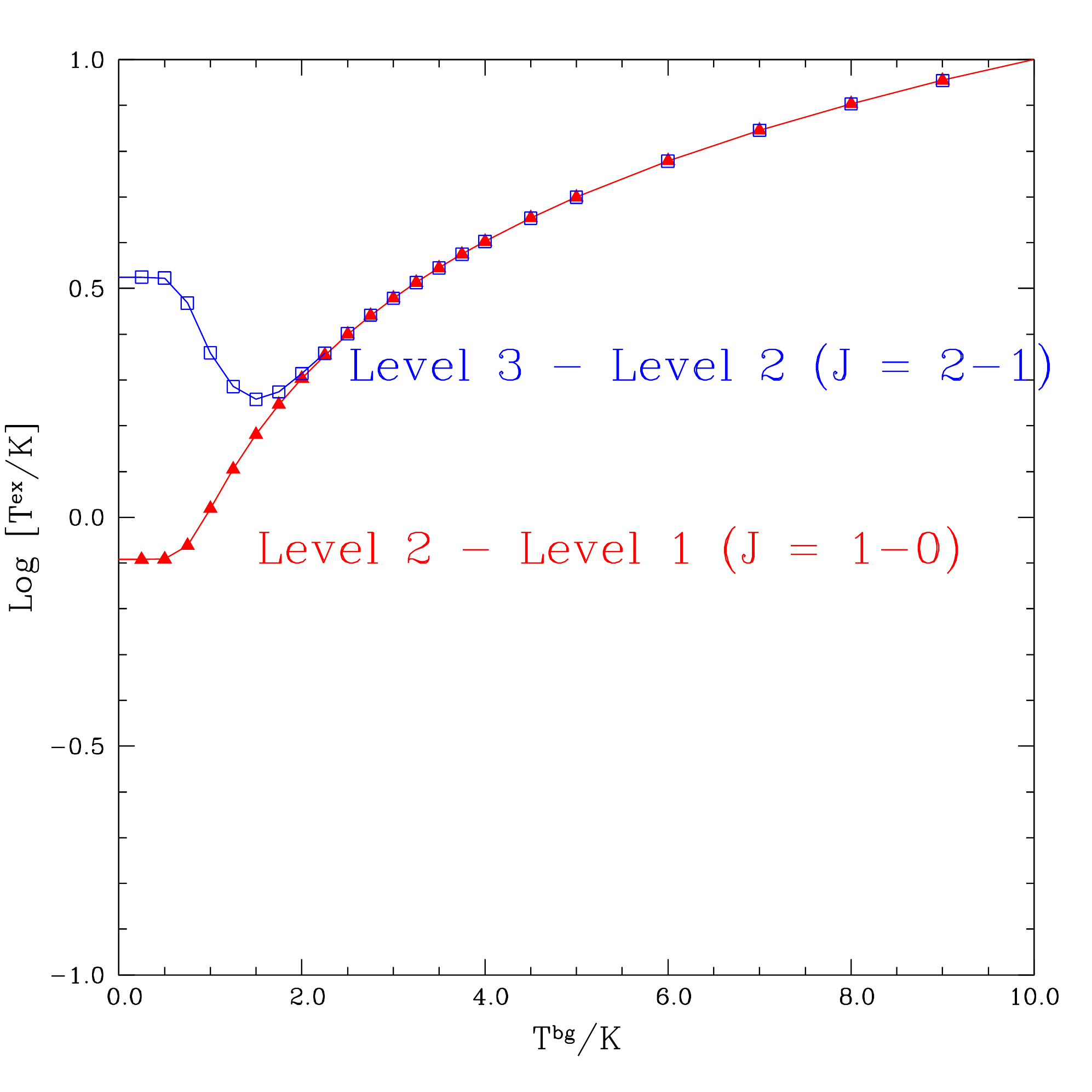}
\caption{\label{tex_tbg} Excitation temperature for two lowest transitions of CO (labeled by rotational quantum number) as a function of the background radiation temperature $T^{bg}$.  The kinetic temperature is 100 K and the \hh\ density is 1 \cc.  
}
\end{center}
\end{figure}
%%%%%%%%%%%%%%%%%%%%%%%%%%%%%%%%%%%%%%%%%%%%%%%

Equation \ref{bg_eq_A} gives $T^{bg'}$ = 0.7 K for the lower transition and 1.2 K for the higher transition, both of which are in reasonable agreement with Figure \ref{tex_tbg}. 
Since $T^{ex}$ for the lower transition is less than $T^{bg'}$, the excitation temperature increases as $T^{bg}$ increases.  
As discussed previously, the excitation temperature of the higher transition is relatively large with no background present, and so it initially drops as $T_{bg}$ increases, before joining the $T^{ex} = T^{bg}$ curve.

For this density and the two lower transitions, $T^{bg'}$ is significantly smaller than the background temperature required to make $B_{ul}U = A_{ul}$, which is just $T^{bg''} = T^*/ln(2) = 1.44T^*$.  
We can see from the excitation temperatures given in Table \ref{radex_co} that the stimulated rate produced by a background temperature equal to 2.7 K is sufficient to bring the excitation temperatures of two lowest transitions close to equilibrium with the the background temperature.
For the higher transitions, the blackbody radiation function falls off sufficiently rapidly that the background becomes insignificant, and the rise of the excitation temperature as one moves up the ladder is essentially the same as from that with no background present at all.

\subsection{Systems with More than Three Levels}
\label{multilevel}

Analytic solution of the level populations in systems having many levels is in general tedious.  
The  exception is for dipole--like collisions for which only adjacent levels are coupled.  
In this situation the ratio of column densities of any pair of adjacent levels can be written

\be
\label{nul_dipole}
\frac{N_u}{N_l} = \frac{B_{lu}U + C_{lu}}{A_{ul} + B_{ul}U + C_{ul}} \lc
\ee
analogous to equation \ref{n32_dipole} and \ref{n21_dipole}, but in which the background radiation can be included.

If collisions connect non--adjacent levels, one typically resorts to numerical solutions based on matrix inversion.  
In the low collision rate limit, the population (for no background radiation) will be limited to the lowest level, since the collisional excitation rate is by assumption less than any spontaneous decay rate.  
In the case of low but nonzero collision rate, the population will be restricted to the lowest few levels.
This will also be the case if there is a background radiation field that produces a stimulated rate comparable to the collision rate for only the lowest few transitions.
For a molecule with simple rotor structure, the analytic solution for no background radiation yields an equation similar to
equation \ref{n32_lowcoll_nobg}, but with some modifications due to the collisions that change the rotational quantum number by a range of integers.  

We can write the population ratio of pair of levels $u$ and $l$ in the absence of any background radiation as
\be
\label{nul_nobg_multilevel}
\frac{N_u}{N_l} = \frac{A_{l~l-1}}{A_{ul}}  \frac{\sum_{k=u}^{k=k_{max}}C_{1k}}{\sum_{k=l}^{k=k_{max}} C_{1k}}
\ee
where $l-1$ indicates the level below the lower level of the pair in question, and $k_{max}$ is the highest level that is connected by collisions to level 1 or the highest level included in the calculation. 
The limits on the summation in the numerator reflect the fact that collisions from the lowest level to levels above the upper level of the pair of interest all decay radiatively much faster than any collisional process, and thus effectively populate the upper level of the pair.  
The sum in the denominator yields the total rate of collisions that populate both members of the pair of levels of interest.

Equation \ref{nul_nobg_multilevel} can be used to obtain the expression for the excitation temperature
\be
\label{tex_nobg_multilevel}
T^{ex}_{ul} = T^*_{ul}/ln[\frac{g_u}{g_l}\frac{A_{ul}}{A_{l~l-1}}P_{ul}] \lc
\ee
where the $P_{ul}$ term reflects the collisional population rates and can be written
\be
P_{ul} = 1 + {C_{1l}}/{\sum_{k=u}^{k=k_{max}} C_{1k}} \lp
\ee
The fractional population of the higher levels will be very small in this limit, but as seen from Table \ref{radex_co}, the excitation temperatures are well--defined.  
It may seem surprising that the higher levels in this case are populated by collisions directly from the lowest level (or levels).  
We can verify this numerically, and for simplicity set the background radiation temperature to zero.  
We use the collision rate coefficients for ortho--\hh\ CO collisions from \citet{flower2001} and \citet{wernli2006} as extrapolated in the LAMDA database (home.strw.leidenuniv.nl/~moldata/), and for definitiveness consider CO rotational levels 20 and 19.  
The collisional deexcitation rate coefficients are $R_{20~1}$ = 7.5$\times$10$^{-17}$ cm$^3$s$^{-1}$ and $R_{20~19}$ = 1.1$\times$10$^{-10}$ cm$^3$s$^{-1}$.
At a kinetic temperature of 100 K, the excitation rates are $R_{1~20}$ = 2.7$\times$10$^{-20}$ cm$^3$s$^{-1}$ and $R_{19~20}$ = 3.9$\times$10$^{-11}$ cm$^3$s$^{-1}$.
The ratio of the rates of population of level 20 from level 1 to that from level 19 is given by
\be
\label{population_rates_ratio}
\frac{\rm{population~rate~from~level~1}}{\rm{population~rate~from~level~19}} = \frac{N_{1}R_{1~20}}{N_{19}~R_{19~20}} \lc
\ee
which in the present case is equal to 2$\times$10$^8$.
It is thus clear that collisions that transfer population from the lowest level (or few lowest levels) to high--lying levels are the dominant excitation mechanism for the higher rotational levels of CO in the low density limit.

Returning to the issue of the expected excitation temperatures for high--J transitions, we must evaluate equation \ref{tex_nobg_multilevel}. 
There are two factors that must be considered to estimate the sum of the collisions to the levels above the upper level of the transition of interest.  
First, the collisional deexcitation rates decrease as $\Delta J$ increases.
Second, the upwards rates (from the ground state) are further reduced by the increasing upper level energy, even for a moderately high kinetic temperature of 100 K.  
The result is that the rate to the lower level of the transition is significantly larger than that  to the upper and to higher--lying levels.

If we consider the transition between levels 10 and 9 ($J$ = 9--8) with ortho--\hh\ collisions  (as discussed above) at 100K, we find $P_{10~9}$ = 3.16, which results in $T^{ex}_{10~9}$ = 30.7 K.
This agrees (fortuitously) well with the 30.6 K Radex result given in Table \ref{radex_co}.
For the transition between levels 18 and 17 ($J$ = 17 -- 16), $P_{18~17}$ = 5.5, which results in $T^{ex}_{18~17}$ = 48.4 K.  
This compares to the RADEX result    $T^{ex}_{18~17}$ = 49.6 K (Table \ref{radex_co}).  
Given we included excitation only up to level 20, the agreement is very satisfactory.

\subsection{Conclusions}

We have analyzed the initially surprising behavior of the excitation of the CO rotational ladder under conditions of very low density, for which the excitation temperature increases steadily as one moves from lower to higher levels.
The same effect is generally observed for rigid rotors, for simple atomic fine structure systems, as well as for molecules with more complex term schemes.
This behavior can be understood by considering the limit in which collisional deexcitation can be ignored.
Radiative decay then makes the population of all levels other than the ground state quite small.  
A (rare) collision from the ground state to an excited state is followed by a radiative cascade.  
The populations of the upper and lower levels of a transition are determined by the collisions into the respective levels, plus the radiative cascade from higher levels.
The result is level populations that depend essentially only on the relative magnitudes of the A--coefficients for decay into  and out of the lower level of the transition of interest. 
In consequence, the excitation temperature is proportional to the equivalent temperature $T^*$ = $hf/k$ of the transition.  
The impact on the lower levels of CO is modest because the stimulated transition rate from the cosmic microwave background radiation is sufficient to make the excitation temperature equal to the background temperature.
The same is not true for the higher levels, for which the background is unimportant.
 
% \twocolumn

%%
%%%%%%%%%%%%%%%%%%%%%%%%%%%%%%%%%%%%%%%%%%%%%%%%%%%%%%%%%%%%
%%
\begin{deluxetable}{lcccc}
\tablecolumns {5}
\tablewidth{0pt}
\tabletypesize{\footnotesize}
\tablecaption{\label{texdata} CO Excitation Temperatures\tablenotemark{1} and \hh\ Column Densities
 from \citet{sheffer2008} Unless Otherwise Indicated}
\tablehead{\colhead{Source} & \colhead{T$^{ex}_{10}$} & \colhead{T$^{ex}_{21}$} & \colhead{T$^{ex}_{32}$} & \colhead{log (N(\hh)/cm$^{-2}$)}}
\startdata
BD +48 3437	&	2.7	& 	&	  &20.42\\
BD +53 2820    &	3.3  &    &      &20.15\\
CPD -69 1743 &  2.7     &     &     &19.99\\
CPD -59 2603 &  3.0  &3.5  &      &20.15\\
HD 12323       &  3.1   &5.3  &      &20.32\\
HD 13268      & 3.4    &        &      &20.51\\
HD 13745      &  4.0   &        &      &20.67\\
HD 14434     &   4.4   &        &      &20.43\\
HD 15137     &   3.1   & 5.1 &      &20.32\\
HD 23478     &   3.4   & 3.7 &5.0 &20.57\\
HD 24190     &   3.1   & 3.7 &      &20.38\\
HD 24398     &   3.4   & 4.0 &5.0 &20.67\\
HD 27778     &   5.3   & 5.6 &5.7 &20.79\tablenotemark{2} \\
HD 30122     &   3.8   & 4.1 &      &20.70\\
HD 36841     &   2.7   & 3.2 &      &20.4\tablenotemark{3}\\
HD 37367     &   3.2   &       &      &20.61\\
HD 37903     &   2.7   &       &      &20.95\\
HD 43818     &   4.1   &       &      &20.4\tablenotemark{3}\\
HD 58510	  &   2.9   &       &      &20.23\\
HD 63005     &   3.6   &       &      &20.23\\
HD 91983     &   2.7   &       &      &20.23\\
HD 93205     &   2.8   &       &      &19.83\\
HD 93222     &   3.3   &       &      &19.84\\
HD 93237     &   3.1   &       &      &19.80\\
HD 93840     &   3.1   &       &      &19.28\\
HD 94454     &   3.8   &       &      &20.76\\
HD 96675     &   3.7   & 8.4 &      &20.86\\
HD 99872     &   3.7   & 3.9 &      &20.52\\
HD 102065   &   3.6   &       &      &20.56\\
HD 106943   &   2.7   &       &      &19.81\\
HD 108002   &   3.2   &       &      &20.34\\
HD 108639   &   3.0   &       &      &20.04\\
HD 110434   &   2.7   &       &      &19.90\\
HD 112999   &  3.0   &       &       &20.11\\
HD 114886   &  3.1   &        &      &20.34\\
HD 115071   &  3.7   &        &      &20.69\\
HD 115455   &  2.9   &        &      &20.58\\
HD 116852   &  3.2   &        &      &19.83\\
HD 122879   &  2.9   &        &      &20.36\\
HD 124314   &  3.4   &        &      &20.52\\
HD 137595   &  3.9   & 4.7  &     &20.62\\
HD 140037   &  2.9   &        &      &19.34\\
HD 144965   &  4.3   & 6.0  &     &20.79\\
HD 147683   &  5.2   & 7.4  &7.4&20.74\\
HD 147888   &  13.6 & 8.0  &8.7&20.58\\
HD 148937   & 3.7    & 4.9  &7.7&20.71\tablenotemark{4}\\
HD 152590   & 4.1    &         &     &20.51\\
HD 152723   & 4.0    &         &     &20.30\\
HD 154368   & 3.0    & 5.5  &     &20.16\tablenotemark{2}\\
HD 157857   & 4.6    &        &     &20.69\\
HD 163758   & 4.0    &        &     &19.85\\
HD 177989   & 3.3    & 3.6  &     &20.15\tablenotemark{4}\\
HD 190918   & 2.7    & 5.3  &     &19.95\\
HD 192035   & 3.2    & 4.4  &     &20.68\\
HD 195965   & 3.0    &        &     &20.34\\
HD 198781   & 3.4    & 3.9 &     &20.56\\
HD  203532  & 5.3    & 4.6  &    &20.70\\
HD  208905   & 6.0    &        &    &20.43\\
HD 209481   & 2.9    &        &     &20.54\\
HD 209975   & 2.9    &        &     &20.15\\
HD 210809   & 3.1    &        &     &20.00\\
HD 220057   & 3.0    &  4.4 &    &20.34\\
HD 303308   & 3.1    &        &    &20.15\\
HD 308813   & 3.8    &        &    &20.30\\
\enddata
\tablenotetext{1} {Excitation temperatures derived from data given in \citet{sheffer2008}, using equations 2 -- 4 in the present paper.}
\tablenotetext{2}{N(\hh) from \citet{rachford2002}.}
\tablenotetext{3} {Estimates of N(\hh) by \citet{sheffer2008} based on correlations with column densities of other species.}
\tablenotetext{4} {N(\hh) from \citet{sheffer2007}.}

\end{deluxetable}

%%
%%%%%%%%%%%%%%%%%%%%%%%%%%%
%%
\begin{deluxetable}{lccccc}
\tablecolumns{6}
\tablewidth{0pt}
\tabletypesize{\footnotesize}
\tablecaption{\label{others} Excitation Temperatures of CO Transitions and \hh\ Column Densities from Other Sources}
\tablehead{\colhead{Source} & \colhead{T$^{ex}_{10}$} & \colhead{T$^{ex}_{21}$} & \colhead{T$^{ex}_{32}$} & \colhead{log (N(\hh)/cm$^{-2}$)} & \colhead{Note}\\
                   \                           & (K)                                  & (K)                                   & (K)                                   &                                                     & }
\startdata
HD24534	&	4.6$\pm$1.7	&5.6$\pm$0.4	&5.6$\pm$1.0	&20.92	&\tablenotemark{a}\\ 
HD104705	&	3.4$\pm$0.8	&				&				&19.98 &\tablenotemark{b}\\  
HD147933	&	2.7$\pm$0.1	&7.6$\pm$0.5	&8.4$\pm$0.5	&20.53 &\tablenotemark{c}\\ 
HD148184	&	3.0$\pm$0.3	&5.2$\pm$0.6	&7.5$\pm$3.7	&20.63 &\tablenotemark{c}\\
HD149757	&	3.4$\pm$0.4	&4.5$\pm$0.2	&5.9$\pm$0.5	&20.62 &\tablenotemark{d}\\
HD185418	&	3.3$\pm$0.3	&4.0$\pm$1.2	&				&20.76	&\tablenotemark{a}\\
HD192639	&	2.6$\pm$1.2	&				&				&20.69 &\tablenotemark{a}\\
HD206267	&	6.3$\pm$1.4	&5.6$\pm$0.5	&6.6$\pm$0.7	&20.86 &\tablenotemark{a}\\
HD207198	&	3.8$\pm$0.5	&3.9$\pm$0.4	& $<$9.1		&20.83	&\tablenotemark{a}\\
HD210121	&	6.2$\pm$3.2	&7.6$\pm$2.4	& 4.2$\pm$0.5	&20.75	&\tablenotemark{a}\\
HD210839	&	3.8$\pm$0.7	&4.2$\pm$0.3	& $<$5.7		&20.84	&\tablenotemark{a}\\
HD218915	&	3.9$\pm$0.2	&				&				&20.15 &\tablenotemark{a}\\
\enddata
\tablenotetext{a}{\citet{sonnentrucker2007}}
\tablenotetext{b}{\citet{burgh2007}}
\tablenotetext{c}{\citet{federman2003} and \citet{savage1977}}
\tablenotetext{d}{\citet{lambert1994} and \citet{savage1977}}
\end{deluxetable}
%%
%%%%%%%%%%%%%%%%%%%%%%%%%%%
%%
%%
%%%%%%%%%%%%%%%%%%%%%%%%%%%
%%
\begin{deluxetable}{ccrcrcc}
\tablecolumns{7}
\tablewidth{0pt}
\tabletypesize{\footnotesize}
\tablecaption{\label{pdrmodels} Results for Models of 2--Sided Slabs}
\tablehead{\colhead{Model} 	&\colhead{Density\tablenotemark{a}}	&\colhead{G\tablenotemark{b}}	&\colhead{Total Extinction $A_v$}	&\colhead{$T_{min}$}	&\colhead{$f_{max}$(\hh)}	&\colhead{$F$(\hh)}\\
										&\colhead{(\cc)}				     			&								     		&\colhead{(mag)}				     	&\colhead{(K)}				&\colhead{Center}				&\colhead{Integrated}
}
\startdata 
1	&50	&1		&1.0	&82	&0.92	&0.79\\
2	&100	&1		&1.0	&62	&0.96	&0.89\\
3	&200	&1		&1.0	&46	&0.98	&0.94\\
4	&100	&10	&1.0	&138	&0.75	&0.46\\
5	&200	&10	&1.0	&122	&0.87	&0.64\\
6	&100	&10	&0.5	&164	&0.37	&0.22\\
7	&100	&10	&2.0	&92	&0.95	&0.71\\
8	&199	&1		&0.2	&106	&0.74	&0.62\\
\enddata
\tablenotetext{a}{$n$(H) = $n$(\ho) + 2$n$(\hh) }
\tablenotetext{b}{Relative to standard interstellar radiation field}
\end{deluxetable}
%%
%%%%%%%%%%%%%%%%%%%%%%%%%%%
%%

\begin{deluxetable}{lcc}
\tablecolumns {3}
\tablewidth{0pt}
\tabletypesize{\footnotesize}
\tablecaption{\label{dens_1-0_only} Limits on Densities for Sources Having Only J = 1--0 Data ($T^k$ = 100 K)}
\tablehead{\colhead{Source} & \colhead{log(n$_{min}$/cm$^{-3}$)} & \colhead{log(n$_{max}$/cm$^{-3}$)}}
\startdata
BD +48 3437	&		&	1.4\\
BD +53 2820	&		&	1.8\\
CPD -69 1743	&		&	1.4\\
HD 13268       	&		&	1.8\\
HD 13745      	&	1.3	&	2.0\\
HD 14434    	&	1.6	&	2.1\\
HD 37367     	&		&	1.7\\
HD 37903     	&		&	1.4\\
HD 43818     	&	1.4	&	2.1\\
HD 58510	  	&		&	1.5\\
HD 63005     	&	0.8	&	1.9\\
HD 91983     	&		&	1.4\\
HD 93205     	&		&	1.4\\
HD 93222     	&		&	1.8\\
HD 93237     	&		&	1.7\\
HD 93840     	&		&	1.7\\
HD 94454     	&	1.2	&	2.0\\
HD 102065   	&	0.8	&	1.9\\
HD 104705	&		&	2.0\\
HD 106943   	&		&	1.4\\
HD 108002   	&		&	1.7\\
HD 108639   	&		&	1.6\\
HD 110434   	&		&	1.4\\
HD 112999	&		&	1.6\\
HD 114886   	&		&	1.7\\
HD 115071   	&	1.1	&	1.9\\
HD 115455   	&		&	1.5\\
HD 116852   	&		&	1.7\\
HD 122879  	&		&	1.5\\
HD 124314   	&		&	1.8\\
HD 140037   	&		&	1.5\\
HD 152590   	&	1.4	&	2.1\\
HD 152723  	&	1.3	&	2.0\\
HD 157857  	&	1.7	&	2.2\\
HD 163758  	&	1.3	&	2.0\\
HD 192639	&		&       1.7\\
HD 195965  	&		&	1.6\\
HD 208905   	&	2.0	&	2.4\\
HD 209481   	&		&	1.5\\
HD 209975   	&		&	1.5\\
HD 210809  	&		&	1.7\\
HD 218915	&	1.6	&	1.9\\
HD 303308   	&		&	1.7\\
HD 308813  	&	1.2	&	2.0
\enddata
\end{deluxetable}

%%%%
\begin{deluxetable}{lcccccccl}
\tablecaption{\label{multiple} Densities\tablenotemark{1} Derived For Sources with Excitation Temperatures for Two or Three Transitions; T$^k$ = 100 K}
\tablecolumns {9}
\tabletypesize{\footnotesize}
%\tabletypesize{\scriptsize}
\tablewidth{0pt}
\tablehead{\colhead{Source}&n$_{min}$&n$_{max}$&n$_{min}$&n$_{max}$&n$_{min}$&n$_{max}$&n$_{min}$\tablenotemark{2}&n$_{max}$\tablenotemark{2}\\
	&\multicolumn{2}{c} {T$^{ex}_{10}$}	&\multicolumn{2}{c} {T$^{ex}_{21}$} &\multicolumn{2}{c} {T$^{ex}_{32}$}&\multicolumn{2}{c} {Combined}}
\startdata

CPD -59 2603 	&	&	1.6	&	0.3	&	1.9	&	-	&	-	&	0.3	&	1.6\\
HD 12323      	&	&	1.7	&	1.9	&	2.3	&	-	&	- 	&	1.8	&	1.8(s)\\
HD 15137     	&	&	1.7	&	1.9	&	2.5	&	-	&	 -	&	1.8	&	1.8(s)\\
HD 23478     	&	&	1.8	&	0.9	&	2.1	&	0.6	&	1.7	&	0.9	&	1.7\\
HD 24190     	&	&	1.7	&	0.9	&	2.1	&	-	&	-	&	0.9	&	1.7\\
HD 24398     	&	&	1.8	&	1.2	&	2.2	&	0.6	&	1.7	&	1.2	&	1.7\\
HD 24534		&0.8 &	2.3	&	2.3	&	2.5	&	0.9	&	2.1	&	2.2	&	2.2(s)\\
HD 27778     	&1.9&	2.3	&	2.1	&	2.6	&	0.7	& 	2.3	&	2.1	& 	2.3 \\
HD 30122     	&1.2&	2.0	&	1.4	&	2.2	&	-	&	-	&	1.4	&	2.0\\
HD 36841     	&	&	1.4	&		&	1.8	&	-	&	-	&		&	1.4\\
HD 96675     	&1.1&	1.9	&	2.6	&	3.0	&	-	&	-	&		&		\\
HD 99872     	&1.1&	1.9	&	1.1	&	2.2	&	-	&	-	&	1.1	&	1.9\\
HD 137595   	&1.2&	2.0	&	1.7	&	2.4	&	-	&	-	&	1.7	&	2.0\\
HD 144965   	&1.5&	2.1	&	2.2	&	2.7	&	-	&	-	&	2.1	&	2.2(s)\\
HD 147683   	&1.9&	2.3	&	2.5	&	2.9	&	1.6	&	2.8	&	2.4	&	2.4(s) \\
HD 147888   	&2.5&	2.7	&	2.6	&	3.0	&	2.3	&	3.0	&	2.6	&	2.7\\
HD 147933		& 	&	0.6	&	2.7	&	2.8	&	2.6	&	2.7	&		&		\\
HD 148184		&	&	1.4	&	2.1	&	2.4	&	0.5	&	3.1	&		&		\\
HD 148937   	&1.1&	1.9	&	1.8	&	2.5	&	1.8	&	2.9	&	1.8	&	1.9\\
HD 149757		&1.1&	1.8	&	2.0	&	2.2	&	1.4	&	2.0	&	1.9	&	1.9(s)\\
HD 154368  	&	&	1.6	&	2.0	&	2.6	&	-	&	-	&		&		\\
HD 177989   	&	&	1.8	&		&	2.0	&	-	&	-	&		&	1.8\\
HD 185418		&1.1	&	1.6	&	0.5	&	2.3	&	-	&	-	&	1.1	&	1.6\\
HD 190918  	&	&	1.4	&	1.9	&	2.6	&	-	&	-	&		&		\\
HD 192035   	&	&	1.7	&	1.5	&	2.3	&	-	&	-	&	1.5	&	1.7\\
HD 198781   	&	&	1.8	&	1.1	&	2.2	&	-	&	-	&	1.1	&	1.8\\
HD  203532 	&1.9&	2.3	&	1.7	&	2.0	&	-	&	-	&	1.9	&	2.0\\
HD 206267		&2.0	&	2.4	&	2.3	&	2.5	&	1.6	&	2.4	&	2.3	&	2.4\\
HD 207198		&1.4	& 	1.9	&	1.5	&	2.0	&		&	2.9	&	1.5	&	1.9\\
HD 210121		&1.1	&	2.5	&	2.3	&	3.0	&	0.4	&	1.0	&		&		\\
HD 210839		&1.2	&	1.9	&	1.8	&	2.1	&		&	1.6	&	1.7	&	1.7(s)\\
HD 220057   	&	&	1.6	&	2.1	&	2.3	&	-	&	-	&		&		\\
\enddata
\tablenotetext{1} {Densities expressed as log(n(H$_2$))}
\tablenotetext{2}{(s) denotes additional 0.1 dex range in assessing consistency among various excitation temperatures}
\end{deluxetable}
%%
%%%%%%%%%%%%%%%%
%%
%%
%%%%%%%%%%%%%%%%
%%
\begin{deluxetable}{lrccc}
%\tabletypesize{\scriptsize}
\tablewidth{0pt}
\tablecaption{\label{average_densities} Average Densities of Different Cloud Categories}
\tablehead{\colhead{Category}	&\colhead{$T^k$}	&\colhead{$<n_{min}>$}	&\colhead{$<n_{max}>$}		&\colhead{$<n_{mid}>$}\\
			                                   &\colhead{(K)}		&\colhead{(\cc)}				&\colhead{(\cc)}				&\colhead{(\cc)}}
\startdata
sources with $T^{ex}_{10}$ only 				&100		&22				&105						&49\\
\\
sources with $T^{ex}_{10}$ \& 	$T^{ex}_{21}$ 	&100		&25				&69						&42\\
sources with $T^{ex}_{10}$ \& 	$T^{ex}_{21}$	&50		&32				&143						&67\\
\\
sources with 3 $T^{ex}$						&100		&75				&118						&94\\
sources with 3 $T^{ex}$						&50		&104				&176						&135\\
\\
sources with 2 or 3 $T^{ex}$					&100		&37				&125						&68\\
sources with 2 or 3 $T^{ex}$					&50		&58				&148						&92\\

\enddata
\end{deluxetable}
%%%%%%%%%%%%%%%%%%
%%%%%%%%%%%%%%%%%%
\begin{deluxetable}{ccccc}
%\tabletypesize{\scriptsize}
\tabletypesize{\footnotesize}
\tablewidth{0pt}
\tablecaption{\label{radex_co}Excitation Temperature of Lower CO Transitions for T$^k$ = 100 K, and n(\hh) = 0.01 \cc\ From RADEX \citep{vandertak2007}}
\tablehead{
\colhead{Upper Level} &\colhead{Lower Level} & \colhead {Frequency}   &\multicolumn{2}{c}{Excitation Temperature T$^{ex}$ (K)} \\
\colhead{}                    &\colhead {}                    & \colhead {(GHz)}          &\colhead{T$^{bg}$ = 0.0 K} &\colhead {T$^{bg}$ = 2.7 K}
}

\startdata
2	&1		& 115.271		&0.53		& 2.70\\
3 	&2		& 230.538		&3.57		& 2.70\\
4	&3		& 345.796		&6.11		& 2.71\\
5	&4		&461.041		&12.08	& 4.01\\
6	&5		&576.268		&15.19	&15.05\\
7	&6		&691.473		&19.01	&19.25\\
8	&7		&806.652		&23.54	&23.27\\
9	&8		&921.800		&27.05	&27.52\\
10	&9		&1036.912	&30.78	&30.58\\
11	&10	&1151.986	&36.51	&35.97\\
12	&11	&1267.015	&36.90	&37.69\\
13	&12	&1381.995	&44.40	&42.74\\
14	&13 	&1496.923	&40.78	&42.80\\
15	&14	&1611.794	&50.24	&48.01\\
16	&15 	&1726.603	&45.42	&47.24\\
17	&16 	&1841.346	&52.27	&50.69\\
18	&17 	&1956.018	&49.58	&51.07\\
19	&18 	&2070.616	&53.60	&53.17\\
20	&19	&2185.135	&51.45	&52.19\\
21	&20	&2299.570	&58.70	&57.95\\
\enddata
\end{deluxetable}

%%%%%%%%%%%%%%%%%%%%%%%%%%%%%%%%%%%%%%%%%%%%%%%%%%%%%%%%%%%%
\end{document}